\documentclass[aps, prl, reprint, superscriptaddress]{revtex4-1}
\usepackage[utf8]{inputenc}
\usepackage{amsmath}
\usepackage{amssymb}
\usepackage{hyperref}
\usepackage{graphicx}
\usepackage{epstopdf}
\usepackage{dcolumn}
\usepackage{bm}
\usepackage{soul}
\usepackage{xspace}
\usepackage{verbatim}
\usepackage{color}

\hypersetup{colorlinks=true,linkcolor=blue,citecolor=blue, filecolor=blue,urlcolor=blue,breaklinks=true}

\usepackage{tikz}
\usetikzlibrary{arrows}

\begin{document}

\title{Sequential measurements for quantum-enhanced magnetometry in spin chain probes}

\author{Victor Montenegro}
\email{vmontenegro@uestc.edu.cn}
\affiliation{Institute of Fundamental and Frontier Sciences, University of Electronic Science and Technology of China, Chengdu 610051, China}

\author{Gareth Si\^{o}n Jones}
\email{gareth.jones.16@ucl.ac.uk}
\affiliation{Department of Physics and Astronomy University College London, Gower Street, London, WC1E 6BT, U.K.}

\author{Sougato Bose}
\email{s.bose@ucl.ac.uk}
\affiliation{Department of Physics and Astronomy University College London, Gower Street, London, WC1E 6BT, U.K.}

\author{Abolfazl Bayat}
\email{abolfazl.bayat@uestc.edu.cn}
\affiliation{Institute of Fundamental and Frontier Sciences, University of Electronic Science and Technology of China, Chengdu 610051, China}

\date{\today}

\begin{abstract}
Quantum sensors outperform their classical counterparts in their estimation precision, given the same amount of resources. So far, quantum-enhanced sensitivity has been achieved by exploiting the superposition principle. This enhancement has been obtained for particular forms of entangled states, adaptive measurement basis change, critical many-body systems, and steady-state of periodically driven systems. Here, we introduce a different approach to obtain quantum-enhanced sensitivity in a many-body probe through utilizing the nature of quantum measurement and its subsequent wave-function collapse without demanding prior entanglement. Our protocol consists of a sequence of local measurements, without re-initialization, performed regularly during the evolution of a many-body probe. As the number of sequences increases, the sensing precision is enhanced beyond the standard limit, reaching the Heisenberg bound asymptotically. The benefits of the protocol are multi-fold as it uses a product initial state and avoids complex initialization (e.g. prior entangled states or critical ground states) and allows for remote quantum sensing.
\end{abstract}

\maketitle

\textit{Introduction.---} Quantum sensing as a key application of quantum technologies~\cite{Degen, braun2018quantum} is now available in various physical setups, including photonic devices~\cite{Pirandola2018, mitchell2004super,nagata2007beating,taylor2013biological,hou2019control, Liu2021}, nitrogen-vacancy centers~\cite{Taylor2008, Bonato2015, blok2014manipulating}, ion traps~\cite{leibfried2004toward,biercuk2009optimized,maiwald2009stylus,baumgart2016ultrasensitive,bohnet2016quantum}, superconducting qubits~\cite{bylander2011noise,bal2012ultrasensitive,yan2013rotating,wang2019heisenberg}, cavity optomechanics~\cite{montenegro2022probing, PhysRevResearch.2.043338, bagci2014optical, Qvarfort2018, PhysRevResearch.3.013159}, and cold atoms~\cite{appel2009mesoscopic,leroux2010implementation,louchet2010entanglement,sewell2012magnetic,bohnet2014reduced,hosten2016measurement}. The precision for estimating an unknown parameter, quantified by the standard deviation $\sigma$, is bounded by the Cram\'{e}r-Rao inequality, $\sigma{\ge}1{/}\sqrt{M\mathcal{F}}$, where $M$ is the number of trials, and $\mathcal{F}$ is the Fisher Information~\cite{Paris, Liu_2019}. For any resource $T$ (e.g., time~\cite{Cappellaro2012,Nusran,Waldherr2011, Said} or number of particles~\cite{Giovannetti2004,Giovannetti2006,Giovannetti2011}), Fisher information, in general, scales as $\mathcal{F}{\sim} T^\eta$. While classical sensors at best results in $\eta{=}1$ (standard limit), quantum sensors can achieve an enhanced sensitivity with $\eta{=}2$ (Heisenberg limit)~\cite{Giovannetti2004, Giovannetti2006, Giovannetti2011}, or even $\eta{>}2$ (super-Heisenberg limit)~\cite{Garbe2021}. A fundamental open problem is to determine which quantum features can be exploited to achieve quantum-enhanced sensing?

Quantum mechanics is distinct from classical physics by two main features, namely quantum superposition and quantum measurements. So far, the superposition principle has been exploited for achieving quantum-enhanced sensitivity through: (i) exploiting the Greenberger-Horne-Zeilinger (GHZ) entangled states~\cite{PhysRevLett.106.110402, Demkowicz2012-nat, PhysRevA.97.042112, PhysRevLett.122.040503, Giovannetti2004, Giovannetti2006, Giovannetti2011, greenberger1989going}; (ii) the ground state of many-body systems at the phase transition point~\cite{PhysRevE.74.031123, PhysRevA.75.032109, PhysRevA.78.042105, PhysRevA.78.042106, FIDELITY2010, skotiniotis2015quantum, PhysRevLett.126.010502, Liu2021, PhysRevLett.126.200501, sarkar2022freefermionic, PhysRevX.8.021022, PhysRevLett.124.120504, PhysRevLett.126.010502}; (iii) the steady-state of Floquet systems~\cite{PhysRevLett.127.080504, mishra2021integrable};  (iv) adaptive~\cite{higgins2007entanglement,Said,Berry2009,Higgins2009,Bonato2015} or continuous measurements~\cite{gammelmark2014fisher,PhysRevLett.125.200505, Albarelli_2017-continuous}; and (v) variational methods for optimizing the initial state as well as the measurement basis~\cite{meyer2021variational,marciniak2021optimal,yang2021variational}. While these methods have their own advantages, they also suffer from several drawbacks. In GHZ-based quantum sensing, the preparation and manipulation are challenging~\cite{kolodynski2013efficient, Demkowicz2012, Albarelli2018restoringheisenberg}, and interaction between particles deteriorates the sensitivity~\cite{boixo2007generalized,de2013quantum,skotiniotis2015quantum,pang2014quantum,de2013quantum}. On the other hand, in both critical and Floquet many-body quantum sensors, the interaction between particles is essential and the system is more robust against imperfections. However, in such quantum sensors the region of quantum-enhanced sensitivity is very narrow~\cite{PhysRevX.8.021022, PhysRevLett.126.200501}. Adaptive measurements are also not practically available in all physical platforms and  training  a  programmable variational quantum sensor may take long times or face convergence issues~\cite{mcclean2018barren}. Projective measurement, as another unique feature of quantum physics, has been employed for quenching many-body systems~\cite{bayat2017scaling, burgarth2014exponential, pouyandeh2014measurement, ma2018phase, PhysRevLett.121.030601} which may induce new types of phase transitions~\cite{PhysRevLett.125.030505, PhysRevLett.127.140601, PhysRevLett.128.010603, PhysRevX.9.031009, PhysRevX.10.041020, PhysRevLett.128.010604}. One may wonder whether projective measurements and their subsequent wave-function collapse, can also be harnessed for obtaining quantum-enhanced sensitivity.
\begin{figure}[t]
\includegraphics[width=\linewidth]{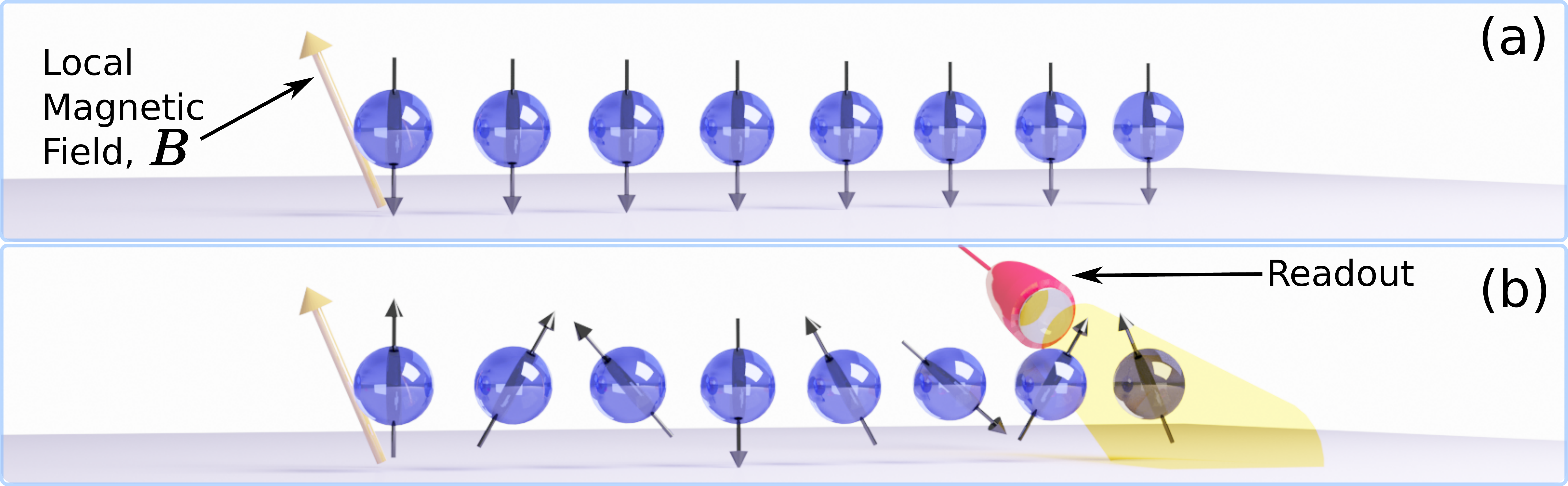}
\caption{Schematic of the protocol. (a) A spin chain probe is initialized in a product state for measuring a local magnetic field $\bm{B}$ at site 1. (b) The readout is performed sequentially on the last site, separated by intervals of free evolution.} \label{fig:the-model}
\end{figure}

Here, we show that quantum measurement and its subsequent wave-function collapse can indeed be used for achieving quantum-enhanced sensitivity. In our proposal, a many-body probe, initialized in a product state, is measured at regular times during its evolution without re-initialization. As the number of subsequent measurements increases, the protocol becomes far more efficient in using time as a resource, and the sensing precision is enhanced beyond the standard limit.

\textit{The Model.---} We consider a spin chain probe made of $N$ interacting spin$-1/2$ particles for sensing a local magnetic field acting upon its first qubit via measuring the last one. For the sake of simplicity, we consider a Heisenberg Hamiltonian:
\begin{equation}
H=-J\sum_{j=1}^{N-1}\boldsymbol{\sigma}_j \cdot \boldsymbol{\sigma}_{j+1} + B_x \sigma_1^x + B_z \sigma_1^z,\label{eq:hamiltonian-single}
\end{equation}
where $\boldsymbol{\sigma}_j{=}(\sigma_j^x{,}\sigma_j^y{,}\sigma_j^z)$ is a vector composed of Pauli matrices acting on qubit site $j$, $J$ is the exchange interaction, and $\bm{B}{=}(B_x{,}0{,}B_z)$ is the local magnetic field to be estimated. While we consider the unknown local field $\bm{B}$ to be in the $xz-$plane, the generalization to the case of $B_y{\neq}0$ is straightforward. The probe is initialized in the ferromagnetic state $|\psi(0)\rangle{=}|\downarrow\downarrow\downarrow\ldots\rangle$ as schematically shown in Fig.~\ref{fig:the-model}(a). Due to the presence of the local magnetic field $\bm{B}$, the system evolves under the action of $H$ as $|\psi(t)\rangle{=}e^{-iHt}|\psi(0)\rangle$. During the evolution, the quantum state accumulates information about the local field $\bm{B}$, which can be inferred through later local measurements on the qubit site $N$, as shown in Fig.~\ref{fig:the-model}(b). As discussed in the Supplemental Material (SM)~\cite{smaterial}, the orientation of the $N$th qubit follows the evolution of qubit 1 with a certain delay. This synchronization allows for remote sensing of $\bm{B}$ by looking at the dynamics at site $N$.

\textit{Sequential Measurement Protocol.---} In a conventional sensing protocol, after each evolution followed by a measurement, the probe is re-initialized, and the procedure is repeated. Typically, initialization is very time-consuming making a significant overhead time for accomplishing the sensing. We propose a profoundly different yet straightforward strategy to use the time resources more efficiently by exploiting measurement-induced dynamics~\cite{bayat2017scaling, burgarth2014exponential, pouyandeh2014measurement, ma2018phase, PhysRevLett.121.030601} and the distinct nature of many-body systems. After initialization, a sequence of $n_\mathrm{seq}$ successive measurements in a single-basis is performed on the readout spin, each separated by intervals of free evolution, without re-initializing the probe. For simplicity, we first focus on the single-parameter estimation, namely $B_z{=}0$, in which $B_x$ is the only parameter to be estimated. In this case, we assume that a simple fixed projective measurement in the $\sigma^z_N$ basis is performed on the last qubit. The steps for data gathering process is then: (i) The system freely evolves according to: $|\psi^{(i)}(\tau_i)\rangle{=}e^{-iH\tau_i}|\psi^{(i)}(0)\rangle$; (ii) The $i$th measurement outcome $|\gamma_i\rangle{=}|{\uparrow}\rangle,|{\downarrow}\rangle$ at site $N$ appears with probability: $p_{\gamma_i}^{(i)}{=}\langle\psi^{(i)}(\tau_i)|\Pi^\gamma_N|\psi^{(i)}(\tau_i)\rangle$, where $\Pi^\uparrow_N{=}(\mathbb{I}{+}\sigma_N^z)/2$ and $\Pi^\downarrow_N{=}(\mathbb{I}{-}\sigma_N^z)/2$ are spin projections; (iii) As a result of obtaining the outcome $\gamma$, the wave-function collapses to the quantum state $|\psi^{(i+1)}(0)\rangle{=}[p_{\gamma_i}^{(i)}]^{-1/2}\Pi^\gamma_N|\psi^{(i)}(\tau_i)\rangle$; and (iv) The new initial state from (iii) is substituted into (i), and the steps are repeated until $n_\mathrm{seq}$ measurements outcomes are consecutively obtained. Note that $|\psi^{(1)}(0)\rangle{=}|\psi(0)\rangle$ is the probe's ferromagnetic initial state, and $\tau_i$ is the evolution time between the $i{-}1$ and $i$ measurements. After gathering a trajectory of length $n_\mathrm{seq}$ of outcomes $\pmb{\gamma}{=}(\gamma_1,\gamma_2,\cdots,\gamma_{n_\mathrm{seq}})$, the probe is reset and the process is repeated to generate a new trajectory. The protocol does not need any prior entanglement as it is built up naturally during the evolution. Due to the entanglement between the readout qubit and the rest of the system, the quantum state of the system after the wave-function collapses still carries information about the local field, which further helps the sensing in the next sequence. Note that, the conventional sensing is a special case of our sequential protocol with $n_\mathrm{seq}{=}1$.

\begin{figure}
\includegraphics[width=\linewidth]{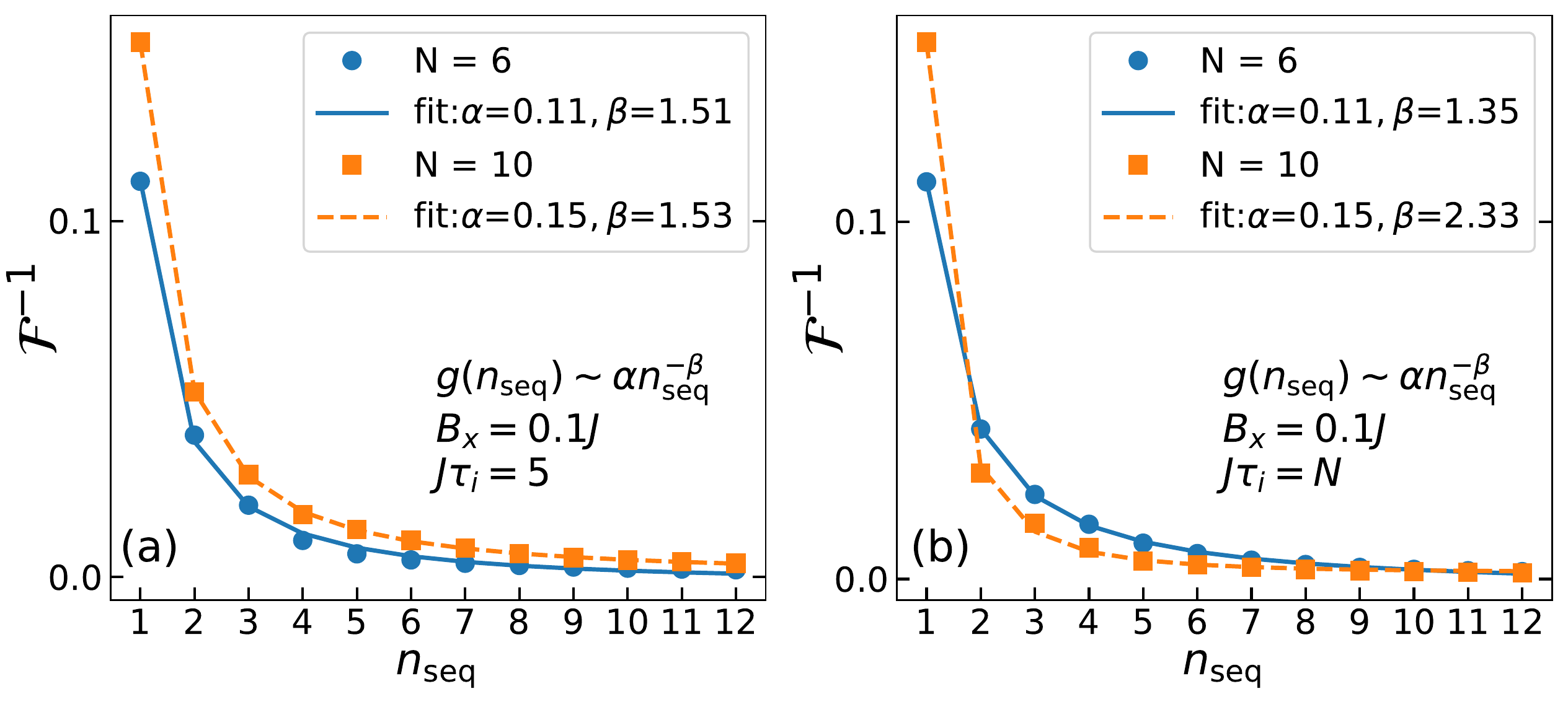}
\caption{Inverse of the Fisher information $\mathcal{F}^{-1}$ as a function of the number of sequential measurements $n_\mathrm{seq}$ performed at site $N$ for $B_x{=}0.1J$. We consider two cases for the time interval between measurements; (a) $J\tau_i{=}5$, and (b) $J\tau_i{=}N$. A fitting function $g(n_\mathrm{seq})$ with exponent $\beta{>}1$ is shown.}
\label{fig:cfi}
\end{figure}

\textit{Classical Precision Bound.---} The sensing precision for estimating $\bm{B}{=}(B_x,0,0)$ given a measurement basis (here $\sigma_N^z$) is determined by the classical Fisher information
\begin{equation}
\mathcal{F}(B_x)=\sum_{\pmb{\gamma}} \frac{1}{P_{\pmb{\gamma}}}\left(\frac{\partial P_{\pmb{\gamma}}}{\partial B_x}\right)^2, \hspace{0.5cm} P_{\pmb{\gamma}}=\prod_{i=1}^{n_\mathrm{seq}}p_{\gamma_i}^{(i)}.
\end{equation}
In the above, $P_{\pmb{\gamma}}$ is the probability of obtaining the trajectory $\pmb{\gamma}$ and the $\sum_{\pmb{\gamma}}$ runs over $2^{n_\mathrm{seq}}$ configurations from $\pmb{\gamma}{=}(\downarrow,\downarrow,\cdots,\downarrow)$ to $\pmb{\gamma}{=}(\uparrow,\uparrow,\cdots,\uparrow)$. To see the impact of sequential measurements on the precision of sensing, in Fig.~\ref{fig:cfi}(a) we plot the inverse of classical Fisher information $\mathcal{F}^{-1}$, as the bound in the Cram\'{e}r-Rao inequality, versus $n_\mathrm{seq}$ for two different probe length $N$ when the unknown parameter $B_x$ is set to be $B_x{=}0.1J$ and $\tau_i{=}\tau{=}5/J$ for all sequences. As the figure clearly shows, $\mathcal{F}^{-1}$ decreases very rapidly by increasing $n_\mathrm{seq}$, indicating the significant advantage of sequential measurements for enhancing the sensing precision. Our numerical data can be fit by $g(n_\mathrm{seq}){=}\alpha n_\mathrm{seq}^{-\beta}{+}\epsilon$ in which $\epsilon$ is vanishingly small and $\beta$ is always found to be $\beta{>}1$. This is indeed an indicator of possible quantum-enhanced sensitivity beyond the standard limit, which will be discussed later. Note that, for larger system sizes, the probe needs more time to transfer information from site 1 to site $N$, and thus, $\tau$ has to be larger. Our numerical investigations show that, $\tau{\sim}N/J$ provides the best estimation. To evidence this, in Fig.~\ref{fig:cfi}(b) we plot $\mathcal{F}^{-1}$ as a function of $n_\mathrm{seq}$ when $\tau_i{=}\tau{=}N/J$ for all sequences. In contrast to Fig.~\ref{fig:cfi}(a), the performance of the longer probe becomes better for this choice of $\tau$. This is an interesting observation, as it shows that using a longer probe facilitates remote sensing and achieves better precision.

\textit{Bayesian Estimation.---} While Fisher information provides a bound for precision, one always needs to use an estimator to actually infer the value of the unknown parameter. Here, we feed the experimental data into a Bayesian estimator, which is known to be optimal for achieving the Cram\'{e}r-Rao bound~\cite{Cramer, Helstrom, Holevo, Braunstein1994, Braunstein1996, Paris, Goldstein, LeCam-1986, Hradil-1996, Pezze-2007, Rubio-2019, Olivares-2009}. By repeating the procedure for $M$ times, one gets a set of $\pmb{\Gamma}{=}\{\pmb{\gamma}_1,\pmb{\gamma}_2,\cdots,\pmb{\gamma}_M\}$, where each trajectory $\pmb{\gamma}_k$ contains $n_\mathrm{seq}$ spin outcomes. Therefore, the total number of measurements performed on the probe is $n_\mathrm{seq}M$. By assuming a uniform prior over the interval of interest, which is assumed to be $B_x{\in}[-0.2J,0.2J]$, one can estimate the posterior distribution $f(B_x|\bm{\Gamma})$. For detailed discussions see the SM~\cite{smaterial}. There are numerous ways to infer $\widehat{B_x}$ as the estimate for $B_x$. Here, we assume that $\widehat{B_x}$ is directly sampled from the posterior distribution $f(B_x|\bm{\Gamma})$. Since $\widehat{B_x}$ is sampled from the probability distribution $f(B_x|\bm{\Gamma})$, one can quantify the quality of the estimation by defining the dimensionless average squared relative error as 
\begin{equation}
\delta B_x^2=\int f(\widehat{B_x}|\bm{\Gamma}) \left(\frac{|\widehat{B_x}-B_x|}{|B_x|}\right)^2d\widehat{B_x},\label{eq:dbx2-simplified}
\end{equation}
where the integration is over the interval of interest, and $|\widehat{B_x}-B_x|/|B_x|$ is the relative error of the estimation. A direct calculation simplifies the above figure of merit as
\begin{equation}
\delta B_x^2=\frac{\sigma^2+|\langle B_x \rangle-B_x|^2}{|B_x|^2},
\end{equation}
where $\sigma^2$ and $\langle B_x \rangle$ are the variance and the average of the magnetic field with respect to the posterior distribution, respectively. The average squared relative error simultaneously quantifies the uncertainty of estimation (i.e., $\sigma$) as well as the bias in the estimation (i.e., $\langle B_x \rangle-B_x$). In the case of unbiased estimator $\delta B_x$ is reduced to $\sigma/|B_x|$ which is the inverse of the signal-to-noise ratio.
\begin{figure}
\includegraphics[width=\linewidth]{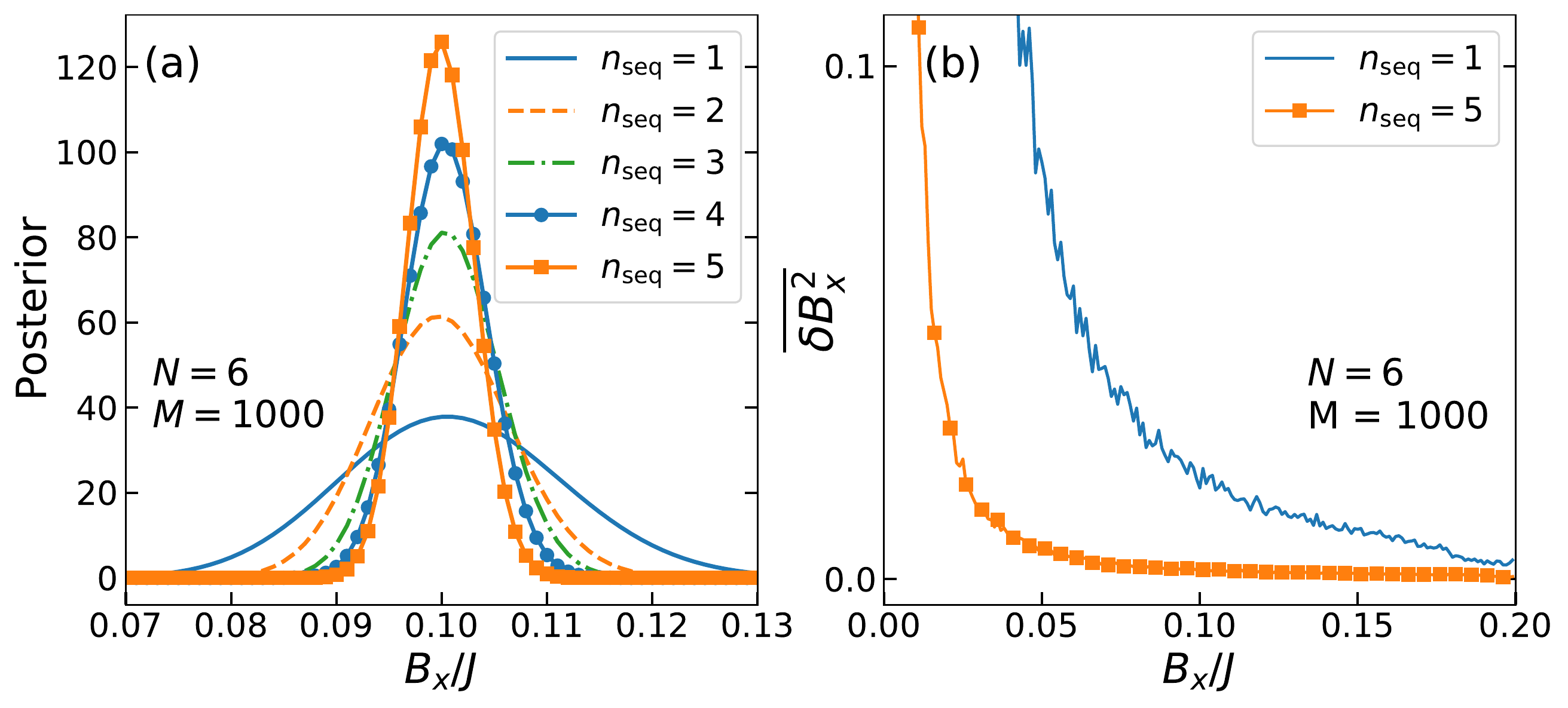}
\caption{(a) Posterior distribution as a function of $B_x/J$ for several $n_\mathrm{seq}$ for sensing $B_x{=}0.1J$. (b) Average of $\delta B_x^2$ as a function of $B_x/J$ for two values of $n_\mathrm{seq}$, where each point is averaged over 100 samples. In both panels, the posterior is obtained by repeating the procedure for $M{=}1000$ times in a probe of $N{=6}$ with fixed $J\tau_i{=}N$.}
\label{fig:posterior-dbx2}
\end{figure}

In Fig.~\ref{fig:posterior-dbx2}(a), we plot the posterior as a function of $B_x/J$ when the actual value is $B_x{=}0.1J$ for different values of $n_\mathrm{seq}$. By increasing the number of sequences, the posterior gets narrower, indicating enhancement of the precision. To show the generality of this across all values of $B_x$, one can compute the average of $\delta B_x^2$ for 100 different samples, denoted by $\overline{\delta B_x^2}$, at each value of $B_x/J$. In Fig.~\ref{fig:posterior-dbx2}(b), we plot $\overline{\delta B_x^2}$ as a function of $B_x/J$ for a probe of length $N{=}6$ and two different values of $n_\mathrm{seq}$. Evidenced by the figure, increasing $n_\mathrm{seq}$ significantly enhances the precision across the whole range of $B_x/J$. Note that, as $B_x/J$ tends to zero, the average error diverges due to the presence of $B_x$ in the denominator of Eq.~\eqref{eq:dbx2-simplified}.

\textit{Trajectory Based Sensing.---} Recently, a numerical analysis~\cite{7403443}, which was followed by analytical proof~\cite{Mael2022}, has shown that by using a single long trajectory $\pmb{\gamma}$ with $n_\mathrm{seq}{\gg}1$ one can reduce the variance of the posterior distribution such that asymptotically reach $f(B_x|\pmb{\gamma}){=}\delta(B_x{-}B_x^\mathrm{real})$, where $B_x^\mathrm{real}$ is the real value of $B_x$ and $\delta(x)$ is the Dirac delta function. This means that a single trajectory with $n_\mathrm{seq}{\gg}1$ is indeed enough to provide an estimation of arbitrary precision. However, it is unclear how the precision scales with $n_\mathrm{seq}$. In what follows, we numerically address this issue.

\begin{figure}[t]
\includegraphics[width=\linewidth]{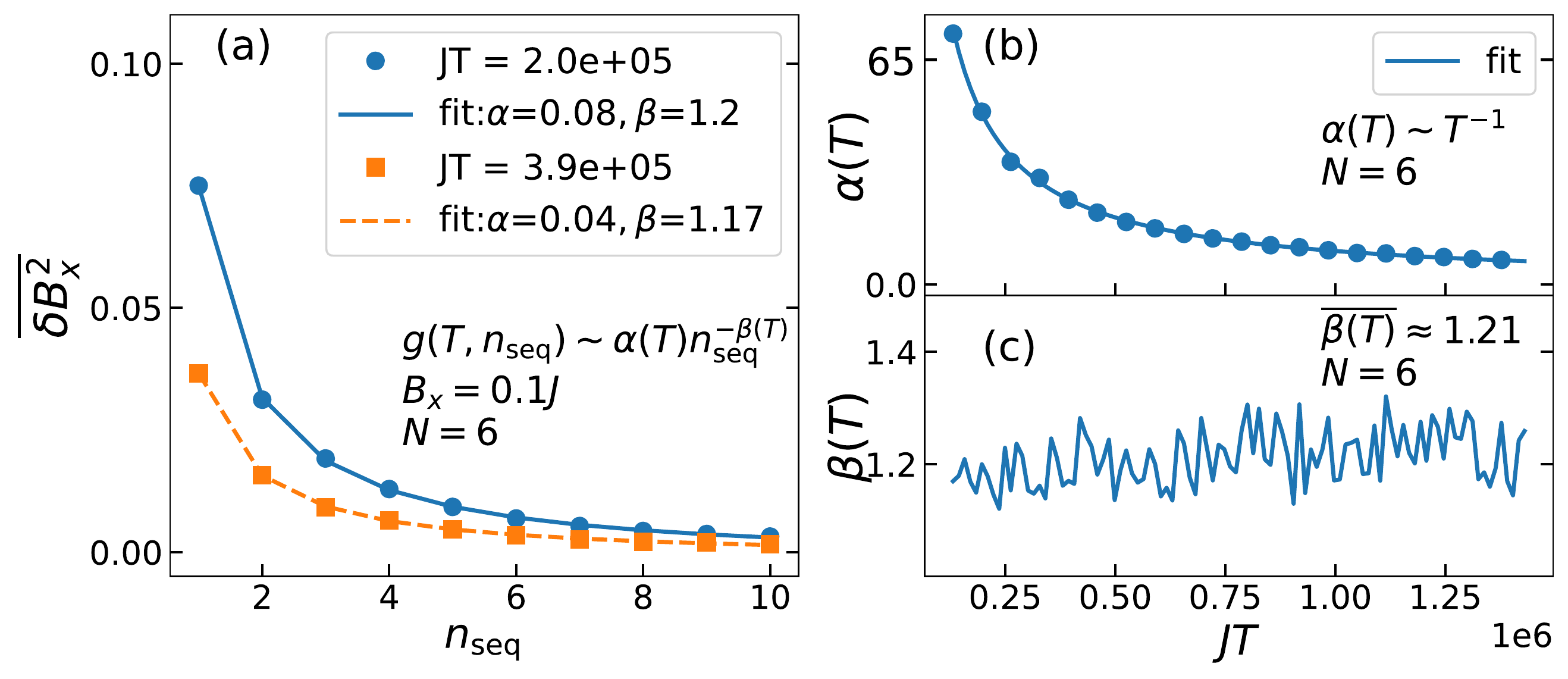}
\includegraphics[width=\linewidth]{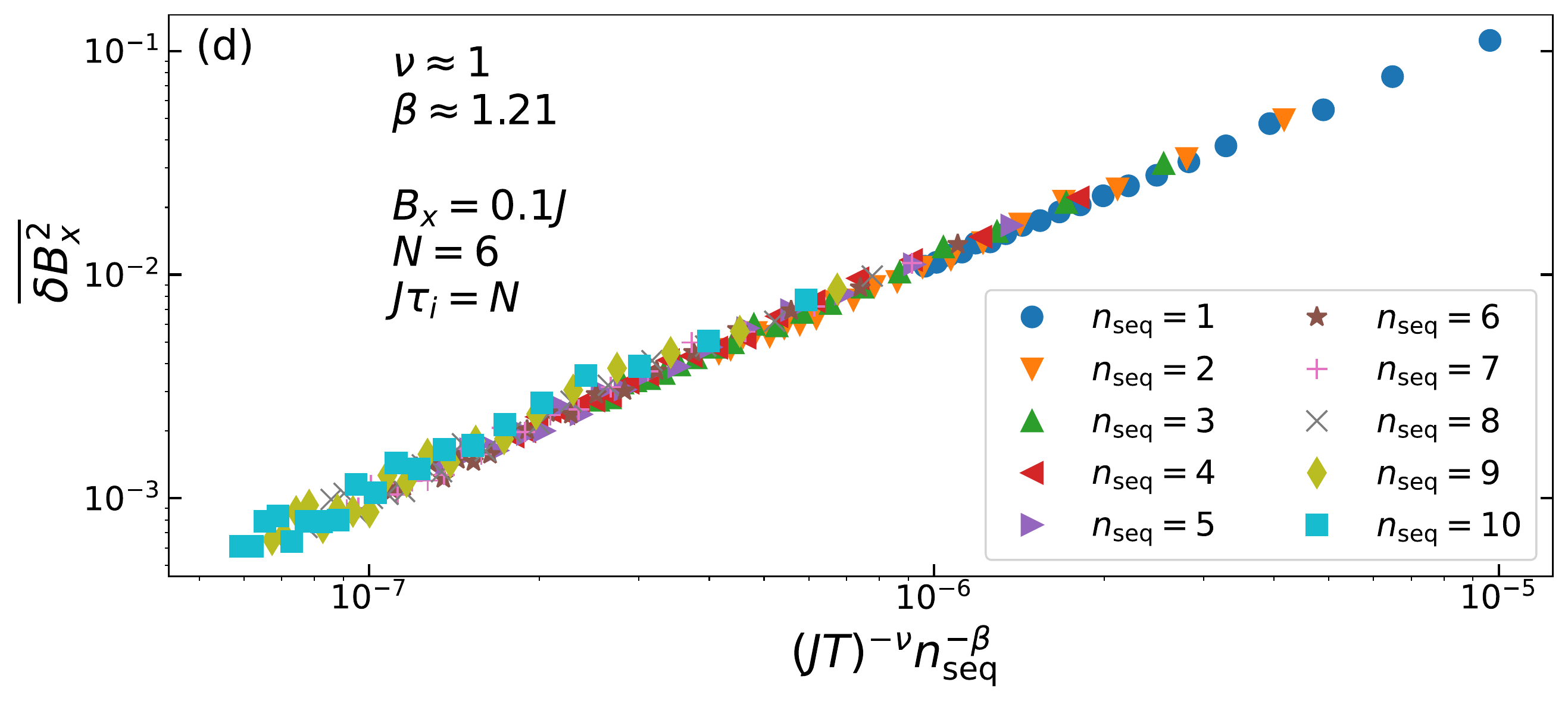}
\caption{Estimating $B_x{=}0.1J$ with a probe of $N{=}6$ and $J\tau_i{=}N$. (a) Averaged squared relative error $\overline{\delta B_x^2}$ versus $n_\mathrm{seq}$ for two total execution times $T$. (b) Fitting coefficient $\alpha(T)$ as function of time $T$. (c) Fitting exponent $\beta(T)$ as function of time. (d) Universal behavior of $\overline{\delta B_x^2}$ versus $(JT)^{-\nu}n_\mathrm{seq}^{-\beta}$ for several values of $n_\mathrm{seq}$ and total time $T$.}
\label{fig:Tn}
\end{figure}

\textit{Time as Resource.---} From a practical point of view, the total time spent for accomplishing the sensing is the main resource to determine the performance of a sensing protocol. While the coherent time evolution of a quantum system is fast, measurement and initialization empirically are one and two orders of magnitude slower, respectively~\cite{Bonato2015}. Therefore, for a given total time, it would be highly beneficial to reduce the number of initialization and use the saved time for increasing the number of measurements. This time compensation allows for a better inference of the information content about the quantity of interest. The total time can be written as 
\begin{equation}
T=M(t_\mathrm{init}+t_\mathrm{evo}+t_\mathrm{meas}n_\mathrm{seq}),\label{eq:total-time}
\end{equation}
where $t_\mathrm{init},t_\mathrm{evo}$, and $t_\mathrm{meas}$ are the initialization, evolution, and measurement times, respectively. By fixing $\tau_i{=}\tau{=}N/J$, one gets $t_\mathrm{evo}{=}n_\mathrm{seq}\tau$. In addition, we fixed $t_\mathrm{init}{=}600/J$ and $t_\mathrm{meas}{=}50/J$, to be consistent with experimental values~\cite{Bonato2015}. For a given total time $T$, the choice of $n_\mathrm{seq}$ changes the re-initialization $M$ and thus the total number of measurements. In Fig.~\ref{fig:Tn}(a), we plot $\overline{\delta B_x^2}$ computed through Bayesian estimation for $B_x{=}0.1J$, as a function of $n_\mathrm{seq}$ for two given values of $T$. Up to a vanishingly small constant, one can use the fitting function $g(T,n_\mathrm{seq}){=}\alpha(T)n_\mathrm{seq}^{-\beta(T)}$. To have a proper resource analysis, one has to determine the dependence of $\alpha(T)$ and $\beta(T)$ exponents with respect to total time $T$. In Figs.~\ref{fig:Tn}(b)-(c), we plot $\alpha(T)$ and $\beta(T)$ as a function of time. While $\alpha(T)$ shows clear dependence on time as $\alpha(T){\sim}T^{-\nu}$, with $\nu{\rightarrow}1$, the exponent $\beta(T)$ fluctuates around 1.21. Thus, the fitting function is reduced to 
\begin{equation}
\overline{\delta B_x^2} \sim T^{-\nu} n_\mathrm{seq}^{-\beta}.\label{eq:dbx2-final-fit}
\end{equation}
This is the main result of our Letter. Note that, although $\nu{\sim}1$ one should not be misled by interpreting it as standard scaling. The key point is that, for a fixed total time $T$ one can always enhance the precision by increasing $n_\mathrm{seq}$. In Fig.~\ref{fig:Tn}(d), we show the universal behavior of Eq.~\eqref{eq:dbx2-final-fit}, through choosing different values of $T$ and $n_\mathrm{seq}$. To better understand the dependence of $\overline{\delta B_x^2} $ on time $T$, one can get $n_\mathrm{seq}{=}(T{-}Mt_\mathrm{init}){/}M(\tau{+}t_\mathrm{meas})$ from Eq.~(\ref{eq:total-time}) and replace it in Eq.~(\ref{eq:dbx2-final-fit}). For $T{\gg}Mt_\mathrm{init}$, one obtains
\begin{equation}\label{eq:asymptotic_M}
	\overline{\delta B_x^2} \sim T^{-(\nu{+}\beta)}.
\end{equation}
As $\nu{\sim}1$ and $\beta{>}1$, one can see that quantum-enhanced sensing can indeed be achieved. Note that the condition $T{\gg}Mt_\mathrm{init}$ can always be satisfied by decreasing the re-initialization $M$ and spending all the time resource $T$ on sequential measurements. In the extreme case of $M{=}1$ ($n_\mathrm{seq}{\gg}1$), one could truly achieve the scaling of Eq.~(\ref{eq:asymptotic_M}). It is also worth  emphasizing that though  $\beta>1$ suggests that our protocol can asymptotically achieve super-Heisenberg scaling (namely $\beta{+}\nu{>}2$), one has to be careful for this generalization in the limit of $n_{seq}{\gg} 1$. Therefore, a more careful investigation remains open for verifying a possible super-Heisenberg precision.

\textit{Protocol Robustness.---} We consider two sources of imperfections, namely dephasing and disordered couplings. As we quantitavely show in the SM~\cite{smaterial}, both of these imperfections are destructive for sensing. Nonetheless, quantum-enhanced sensitivity, i.e., super linear scaling of $\mathcal{F}$, can be found until dephasing or disorder strength are greater than a threshold value. Beyond these threshold values, $\mathcal{F}$ scales linearly with $n_\mathrm{seq}$ and quantum-enhanced sensitivity is lost. Our numerical simulations, see SM~\cite{smaterial}, show that for both dephasing and disorder strengths of up to $\sim$5\%$J$ the quantum-enhanced sensitivity can be achieved.

\begin{figure}[t]
	\includegraphics[width=\linewidth]{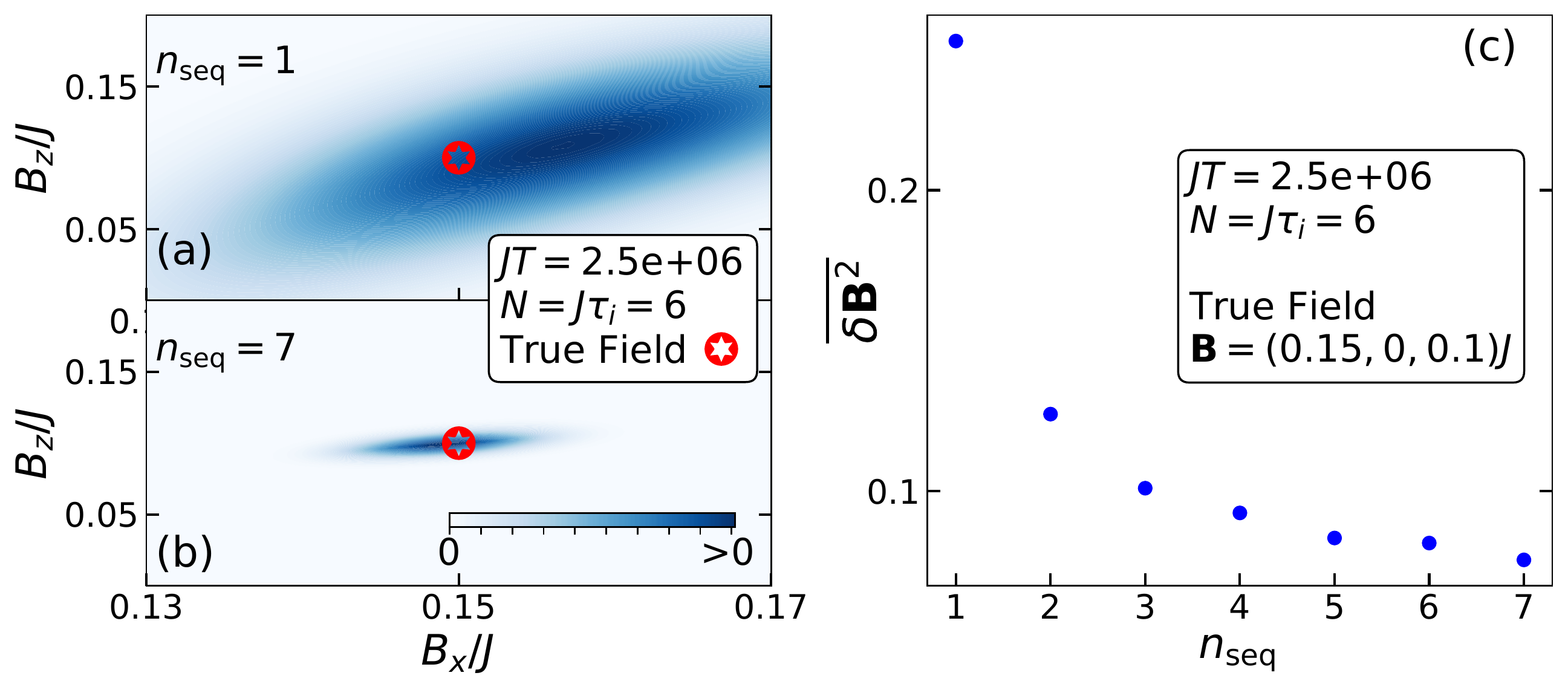}
	\caption{Panels (a) and (b) show the posteriors distributions for $n_\mathrm{seq}{=}1$ and $n_\mathrm{seq}{=}7$ for the estimation of $\bm{B}{=}(0.15,0,0.1)J$, respectively. (c) Averaged squared relative error $\overline{\delta \bm{B}^2}$ as a function of $n_\mathrm{seq}$.}
	\label{fig:gradient-2d}
\end{figure}

\textit{Two parameter estimation.---} For the sake of completeness, we also consider two parameter sensing in which both $B_x$ and $B_z$ are non-zero. In this case, a single $\sigma_N^z$ measurement is not enough to estimate both of the parameters. Hence, we consider a positive operator-valued measure built from the eigenvectors of $\sigma_N^z$ and $\sigma_N^x$~\cite{smaterial}. To exemplify the performance of our protocol, we consider $\bm{B}{=}(0.15,0,0.1)J$, and for a given time $T$ we perform Bayesian analysis for two values of $n_\mathrm{seq}$. In Figs.~\ref{fig:gradient-2d}(a)-(b), we plot the posterior $f(\bm{B}|\pmb{\Gamma})$ in the plane of $B_x/J$ and $B_z/J$ for $n_\mathrm{seq}{=}1$ and $n_\mathrm{seq}{=}7$, respectively. Remarkably, the posterior shrinks significantly as $n_\mathrm{seq}$ increases indicating the effectiveness of sequential measurements for enhancing the precision for a given time. To further clarify this, we can generalize the average squared relative error in Eq.~\eqref{eq:dbx2-simplified} by replacing $B_x$ with $\pmb{B}$ (and $|\cdot|$ represents the norm of the vector) to obtain $\overline{\delta \pmb{B}^2}$. In Fig.~\ref{fig:gradient-2d}(c), we plot $\overline{\delta \pmb{B}^2}$ as a function of $n_\mathrm{seq}$ for $\bm{B}{=}(0.15,0,0.1)J$ which shows rapid enhancement as the number of sequences increases. This clearly shows the generality of our protocol for multi-parameter sensing.

\textit{Conclusions.---} We propose a protocol for remotely sensing a local magnetic field through a sequence of local measurements performed on a single qubit of a quantum many-body probe initialized in a product state. By increasing the sequence of measurements one can avoid the time consuming probe's re-initialization allowing for taking more measurements within the same amount of time. This naturally enhances the sensing precision which asymptotically reaches the Heisenberg bound. Unlike previous schemes, our procedure utilizes the nature of quantum measurement and its subsequent wave-function collapse, and thus, avoids the need of complex initial entangled states, quantum criticality and adaptive measurements. Unlike the protocols based on deferred measurement schemes~\cite{PhysRevLett.98.090401}, our protocol neither requires ancilla qubits nor relies on feedback control. Thus, our minimal control scheme is expected to be less demanding for practical implementations.

\textit{Acknowledgments.---} A.B. acknowledges support from the National Key R\& D Program of China (Grant No. 2018YFA0306703), the National Science Foundation of China (Grants No. 12050410253 and No. 92065115), and the Ministry of Science and Technology of China (Grant No. QNJ2021167001L). V.M. thanks the National Natural Science Foundation of China (Grant No. 12050410251), the Chinese Postdoctoral Science Fund (Grant No. 2018M643435), and the Ministry of Science and Technology of China (Grant No. QNJ2021167004). SB acknowledges the EPSRC grant for nonergodic quantum manipulation (Grant No. EP/R029075/1).

\bibliographystyle{apsrev4-1}

\bibliography{Remote_Sensing}

\clearpage
\onecolumngrid
\pagebreak
\widetext

\begin{center}
\textbf{\large Supplemental Material: Sequential measurements for quantum-enhanced magnetometry in spin chain probes}

\vspace{0.25cm}

Victor Montenegro$^{1}$, Gareth Si\^{o}n Jones$^{2}$, Sougato Bose$^{2}$, and Abolfazl Bayat$^{1}$

\vspace{0.25cm}

$^{1}${\small \em Institute of Fundamental and Frontier Sciences,\\ University of Electronic Science and Technology of China, Chengdu 610051, PR China}\\
$^{2}${\small \em Department of Physics and Astronomy University College London, Gower Street, London, WC1E 6BT, U.K.}

\end{center}

\date{\today}
\setcounter{equation}{0}
\setcounter{figure}{0}
\setcounter{table}{0}
\setcounter{page}{1}
\makeatletter
\renewcommand{\theequation}{S\arabic{equation}}
\renewcommand{\thefigure}{S\arabic{figure}}

The present Supplemental Material clarifies aspects of the remote feature of our sensing protocol, some brief elements on Bayesian estimation, the general case for the estimation of a multi-component local magnetic field, and the robustness of our protocol in the presence of noise and decoherence.

\section{I. sensing at a distance}

In the presence of a non-zero field $\bm{B}$, the initial probe's product state $|\psi(0)\rangle=|\downarrow,\downarrow,\ldots,\downarrow\rangle$ is not an eigenstate of the Hamiltonian, and hence evolves under the action of $H$ (see Eq.~\eqref{eq:hamiltonian-single} in main text). To see how our protocol readily enables remote sensing, for the sake of simplicity, let us consider a local magnetic field only at site 1 in the $x$-direction, i.e., $\bm{B}=(B_x,0,0)$. To evidence the influence of a non-zero local field at site 1 and its corresponding remote effect at the last site, we compute the magnetization in the $z$-direction at qubit site $j$ as follows:
\begin{equation}
m_j(t)=p_\uparrow(t) - p_\downarrow(t)=2 \langle \psi(t)|\Pi_j^\uparrow|\psi(t)\rangle - 1,
\end{equation}
where
\begin{eqnarray}
p_\uparrow(t) &=& \langle \psi(t)|\Pi_j^\uparrow|\psi(t)\rangle,\\
\Pi_j^\uparrow &=& \frac{\mathbb{I}+\sigma_j^z}{2},\\
\Pi_j^\downarrow &=& \frac{\mathbb{I}-\sigma_j^z}{2}.
\end{eqnarray}

\begin{figure}[h!]
\includegraphics[width=0.4\linewidth]{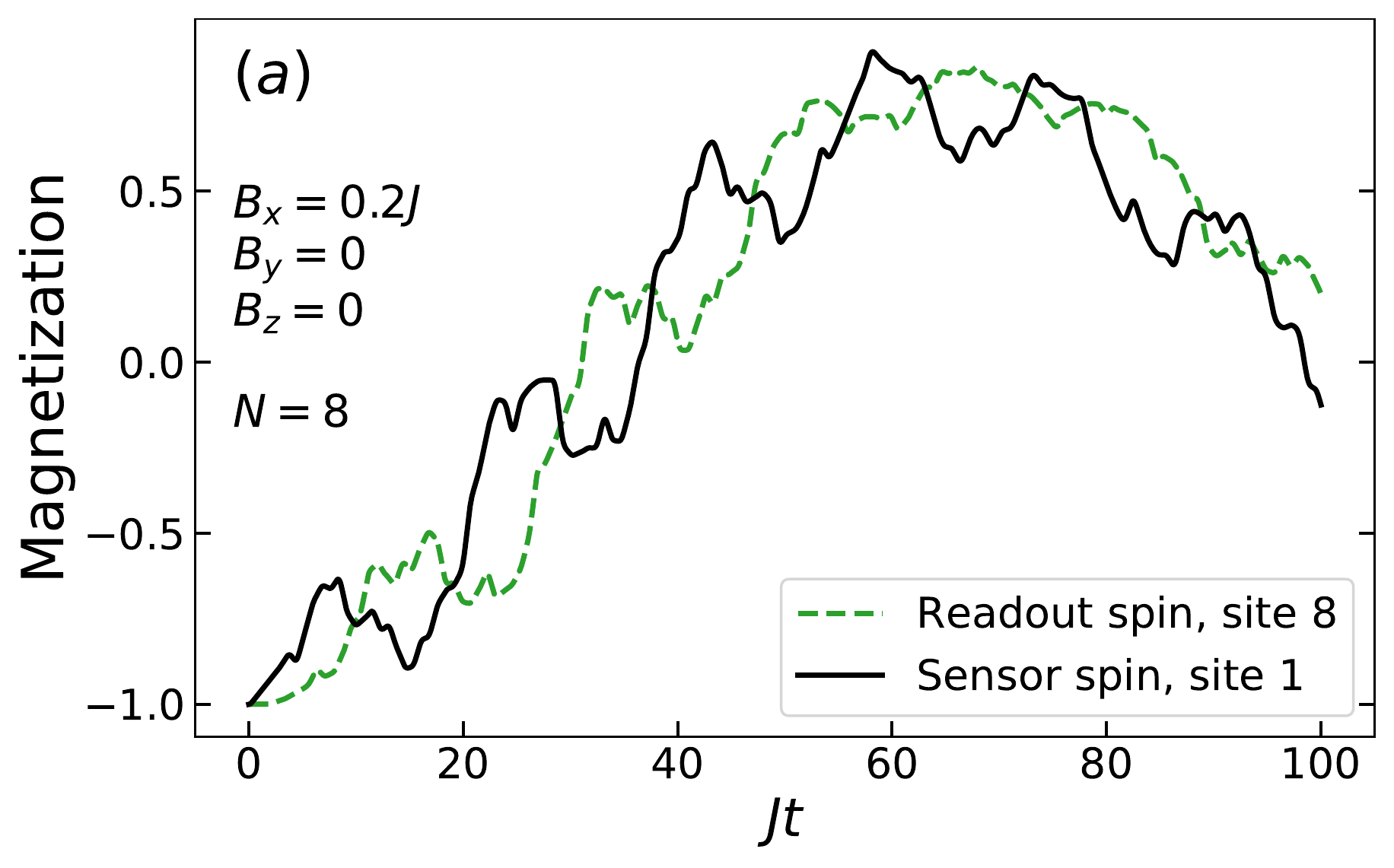}
\includegraphics[width=0.4\linewidth]{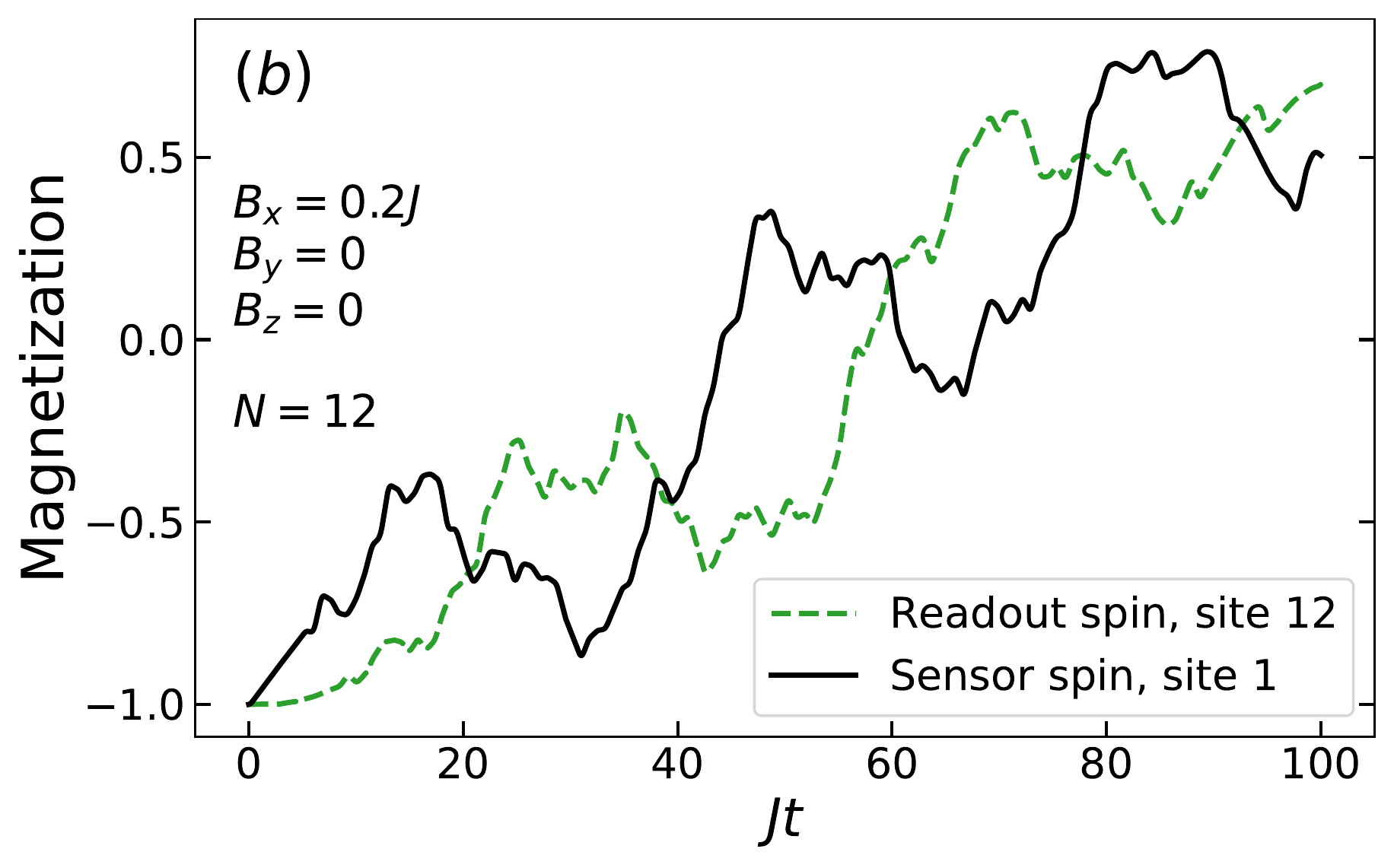}
\caption{The magnetization of the first and last sites as a function of time for $B_x/J=0.2$ for a system size of (a) $N=8$, and (b) $N=12$. The dynamics of both sites are simultaneously affected after some delay.}
\label{fig:magnetization}
\end{figure}

In Figs.~\ref{fig:magnetization}(a)-(b), we depict the magnetization of both the first and last sites as a function of time for a fixed $B_x/J = 0.2$ and two system sizes of length $N = 8$ and $N = 12$, respectively. As seen from the figures, the magnetization at the readout site $N$ evolves in time, roughly synchronizing with the magnetization of the sensor site $m_1(t)$ after a certain delay. This means that by looking at the dynamics at site $N$, one can gain information about the local field $B_x$ remotely. Note that, as evidenced from the figures, the time delay becomes more apparent by increasing the length of the probe. This delay, dictated by the chain's length, can be understood as the needed time for the local magnetic field to transfer the information from site 1 to $N$ through flipping neighboring spins.

\section{ii. Inferring the local magnetic field through Bayesian analysis}

Any sensing procedure for the estimation of an unknown quantity follows three main steps:
\begin{enumerate}
\item Choosing an appropriate probe for encoding the unknown parameter(s),
\item Gathering data for the relevant quantities via measurements on the chosen probe,
\item Processing the gathered data from the last step with an estimator to infer the unknown parameters.
\end{enumerate}
While each step of a sensing protocol must be optimized to achieve the ultimate bound precision, the election of an optimal probe and the optimal measurement(s) are not always practically available. Hence, it is highly desirable to initialize and measure the system with undemanding experimental available resources. These had been specially considered in our sequential protocol as: (i) we initialize the probe in a product state without requiring any initial complex entangled state; and (ii) the local single-site spin measurement is performed on the computational basis. On the other hand, processing the collected data to infer the unknown parameters can be curated through any estimation methodology, for instance, employing statistical frequentist analysis or machine learning-like techniques. In particular, we use Bayesian analysis throughout our work to infer the unknown magnetic field. Notably, it has been reported that processing the gathered data with Bayesian estimators is optimal for achieving the Cram\'{e}r-Rao bound in the limit of large data sets and also shows to be an excellent estimator in the scenario of finite collected data~\cite{Rubio-2019}. In what follows, we briefly describe the Bayesian analysis employed throughout our numerical simulations.

In a nutshell, Bayesian analysis is the art of assigning a probability to a specific event based on updating our knowledge from an educated guess. Bayes' theorem is based on three main statistical concepts, namely:
\begin{enumerate}
\item Conditional probability, $f(A|C)$, which quantifies the probability for the the outcome $A$ to occur given that event $C$ happens ---conditional probability is not interchangeable, i.e., $f(A|C) \neq f(C|A)$,
\item Marginal probability, $f(A)$, which gives the probability of the event $A$ to occur regardless of any other event,
\item Joint probability, $f(A \cap C)$, is the probability of $A$ and $C$ occurring together.
\end{enumerate}
Note that one can write the joint probability in terms of the marginal and conditional probabilities as $f(A \cap C) = f(A|C)f(C)$. Since the joint probability is interchangeable, one can write:
\begin{eqnarray}
 \nonumber f(A \cap C) &=& f(C \cap A),\\
 f(A|C) &=& \frac{f(C|A)f(A)}{f(C)}. \label{eq:bayes-rule}
\end{eqnarray}
The above expression in Eq.~\eqref{eq:bayes-rule} is the Bayes theorem, and it constitutes a sophisticated way of revising our set of beliefs for the occurrence of the event $A$ given that $C$ happens. The denominator $f(C)$ in Eq.~\eqref{eq:bayes-rule}, as the marginal probability of $C$, accounts for a normalization factor such that $f(A|C)$ represents a probability distribution. The marginal probability $f(A)$ is called \textit{prior}, and it accounts for our degree of initial guess of finding the outcome $A$. The procedure of updating our guess via measurements is imprinted in $f(C|A)$, the so-called \textit{likelihood}. The left-hand side in the Bayes' theorem, $f(A|C)$, is called \textit{posterior}. Once the Bayesian analysis is finished, one uses the posterior probability distribution as the new prior, and thus, our set of beliefs is refined.

\textbf{\textit{Example:}} To further illustrate the Bayesian mechanism, we will present the Bayes' rule in practice for a single-qubit model. Let us consider a single-qubit initialized in the state $|+\rangle = (|0\rangle + |1\rangle)/\sqrt{2}$ in the presence of a magnetic field $B$ with Hamiltonian $H = B\sigma_z$. To gather relevant data regarding the magnetic field $B$, we perform measurements on the qubit probe with a chosen measurement basis given by $\{|+\rangle, |-\rangle\}$. The probability of finding the state in $|+\rangle$ after a time $t$ is $p_+ = \cos^2(Bt)$. To build the likelihood, one notices that by repeating $M$ times the procedure, one obtains a random data set of $\{+, -\}$ outcomes corresponding to finding the state in $|+\rangle$ or $|-\rangle$, respectively. The probability distribution of finding $k$ outcomes $+$ out of $M$ trials follows the binomial distribution, and therefore one can readily write the likelihood as follows:
\begin{equation}
f(\text{observed data} \rightarrow (M, k)|B) = \binom{M}{k} p_+^k p_-^{M-k} = \binom{M}{k} \cos^{2k}(Bt) \sin^{2(M-k)}(Bt).\label{eq:singlequbit-likelihood}
\end{equation}
The above likelihood function is essentially the conventional coin tossing example using Bayes' rule, here applied to qubit sampling. The prior, on the other hand, for the first run of the experiment is assumed to be a uniform distribution within some interval that we believe the magnetic field $B$ belongs. This accounts for having no knowledge about the amplitude of the magnetic field but only where it lies within that interval. The posterior is then written simply as:
\begin{equation}
f(B|\text{observed data} \rightarrow (M, k)) \propto \binom{M}{k} \cos^{2k}(Bt) \sin^{2(M-k)}(Bt).
\end{equation}
Note that from the experiment, one only obtains the observed data of finding $k$ outcomes $+$ out of $M$ trials. After the Bayesian rule is finished, we have gained more information regarding the magnetic field $B$. Thus, for the second run of sampling data, one can update the prior (no longer a uniform distribution over the interval where we believe the magnetic field belongs) with the new revisited set of beliefs (i.e., the posterior obtained from the first run of experiments). This is the updating mechanism of the educated guess behind the Bayes' theorem.

We are now in position of explaining in detail the Bayes analysis for our sequential protocol. The Bayes' theorem applied to our work is:
\begin{equation}
f(B_x|\bm{\Gamma})=\frac{f(\bm{\Gamma}|B_x)f(B_x)}{f(\bm{\Gamma})},\label{eq:bayes-theorem-sm}
\end{equation}
where we have considered for the sake of simplicity the single-parameter estimation scenario, i.e. the estimation of local magnetic field of the form $\bm{B}=(B_x,0,0)$. In Eq.~\eqref{eq:bayes-theorem-sm}, $f(B_x)$ is the prior probability distribution for $B_x$, $f(B_x|\bm{\Gamma})$ known as the \textit{posterior} is the probability distribution for the magnetic field $B_x$ given a set of measurement outcomes $\bm{\Gamma}$, $f(\bm{\Gamma}|B_x)$ is the \textit{likelihood} function, and the denominator $f(\bm{\Gamma})$ is a normalization factor such that
\begin{equation}
\int f(B_x'|\bm{\Gamma}) dB_x' = 1.
\end{equation}
Note that one can readily observe the power of the Bayes' rule, the left-hand side of Eq.~\eqref{eq:bayes-theorem-sm}, as one can assign the probability of finding $B_x$ provided a set of observed data $\bm{\Gamma}$ (this is what one would want in actual experiments). In contrast, the opposite scenario would face way more difficulties, i.e., assigning a probability of finding a set of measurements $\bm{\Gamma}$ given an \textit{unknown} magnetic field $B_x$. Thus, one needs to focus on calculating the likelihood $f(\bm{\Gamma}|B_x)$. To compute the likelihood function, $f(\bm{\Gamma}|B_x)$, and therefore the posterior, one needs to repeat the sequential protocol $M$ number of times. After $M$ repetitions of the protocol, one gets a data set $\pmb{\Gamma}{=}\{\pmb{\gamma}_1,\pmb{\gamma}_2,\cdots,\pmb{\gamma}_M\}$ of $M$ trajectories, where each trajectory $\pmb{\gamma}_k$ contains a string of $n_\mathrm{seq}$ spin outcomes, performed at some sequential measurement times $\{\tau_1,\tau_2,{\ldots},\tau_{n_\mathrm{seq}}\}$. Similar to the above single-qubit example, where the probability distribution in Eq.~\eqref{eq:singlequbit-likelihood} is built by \textit{counting} how many $\{+, -\}$ outcomes appear out of $M$ trials, and consequently, following a binomial distribution. Here, the likelihood function $f(\bm{\Gamma}|B_x)$ is built by \textit{counting} how many trajectories $\pmb{\gamma}_j$ appear out of $M$ trials, and therefore, it follows a multinomial distribution as:
\begin{equation}
f(\bm{\Gamma}|B_x)=\frac{M!}{k_1!k_2!\cdots k_{2^{n_\mathrm{seq}}}!}\prod_{j=1}^{2^{n_\mathrm{seq}}}\left[f(\pmb{\gamma}_j|B_x)\right]^{k_j},
\end{equation}
where $k_1,{\cdots},k_{2^{n_\mathrm{seq}}}$ represent the number of times that the trajectory $\pmb{\gamma}_1=(\uparrow_1,\uparrow_2,\ldots,\uparrow_{n_\mathrm{seq}})$ to $\pmb{\gamma}_{2^{n_\mathrm{seq}}}=(\downarrow_1,\downarrow_2,\ldots,\downarrow_{n_\mathrm{seq}})$ occurs in the whole sampling data set $M$ with the constraint $k_1 + k_2 + \cdots + k_{2^{n_\mathrm{seq}}}=M$. The term $f(\pmb{\gamma}_j|B_x)$ accounts for the probability distribution for the trajectory $\pmb{\gamma}_j$ assuming the magnetic field $B_x$, and thus, one requires to classically simulate the probability distributions for all the possible trajectories from $\pmb{\gamma}_1$ to $\pmb{\gamma}_{2^{n_\mathrm{seq}}}$ over a relevant range of $B_x$. To mimic an experimental procedure, one randomly generates a set of $\bm{\Gamma}$ from the corresponding probability distributions with the observed data being the number of occurrences of sequences $\pmb{\gamma}_k$. A generalization of the Bayesian analysis for multi-parameter estimation is found below.

\section{iii. Multi-directional local magnetic field estimation}

As discussed in the main text, our sequential sensing protocol is general and can be employed to estimate a multi-component local magnetic field. To do so, we generalize straightforwardly the Bayes' rule in Eq.~\eqref{eq:bayes-theorem-sm} as
\begin{equation}
f(B_x,B_z|\bm{\Gamma})=\frac{f(\bm{\Gamma}|B_x,B_z)f(B_x,B_z)}{f(\bm{\Gamma})},
\end{equation}
where the denominator imposes 
\begin{equation}
\int f(B_x', B_z'|\bm{\Gamma}) dB_x'dB_z' = 1,
\end{equation}
and the bi-valued likelihood function now reads:
\begin{equation}
f(\bm{\Gamma}|B_x,B_z)=\frac{M!}{k_1!k_2!\cdots k_{2^{n_\mathrm{seq}}}!}\prod_{j=1}^{2^{n_\mathrm{seq}}}\left[f(\pmb{\gamma}_j|B_x,B_z)\right]^{k_j}.
\end{equation}
To compute the probability distribution $f(\pmb{\gamma}_j|B_x,B_z)$ for an outcome measurement sequence $\pmb{\gamma}_j$ assuming the field's components $B_x$ and $B_z$, we consider a positive operator-valued measure (POVM) built from the eigenvectors of $\sigma_N^z$ and $\sigma_N^x$ as: 
\begin{eqnarray}
\Pi_N^\uparrow &=& \frac{\mathbb{I}+\sigma_N^z}{4}, \hspace{1cm} \Pi_N^\downarrow = \frac{\mathbb{I}-\sigma_N^z}{4},\\
\Pi_N^+ &=& \frac{\mathbb{I}+\sigma_N^x}{4}, \hspace{1cm} \Pi_N^- = \frac{\mathbb{I}-\sigma_N^x}{4}.
\end{eqnarray}
The above set of POVMs correspond to measuring half of the times $\sigma_N^z$ and half of the times $\sigma_N^x$. Hence, one can split the $M$ re-initialization sampling of the probe by gathering data $M/2$ times via $\sigma_N^z$ and the rest $M/2$ times via $\sigma_N^x$. 

\begin{figure}
\includegraphics[width=0.75\linewidth]{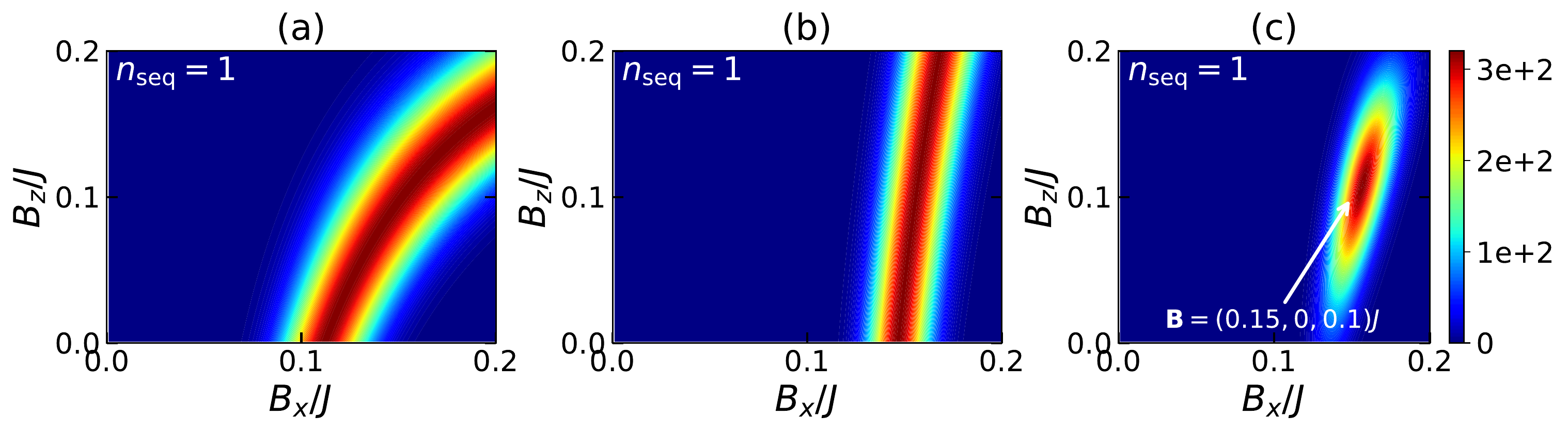}
\includegraphics[width=0.75\linewidth]{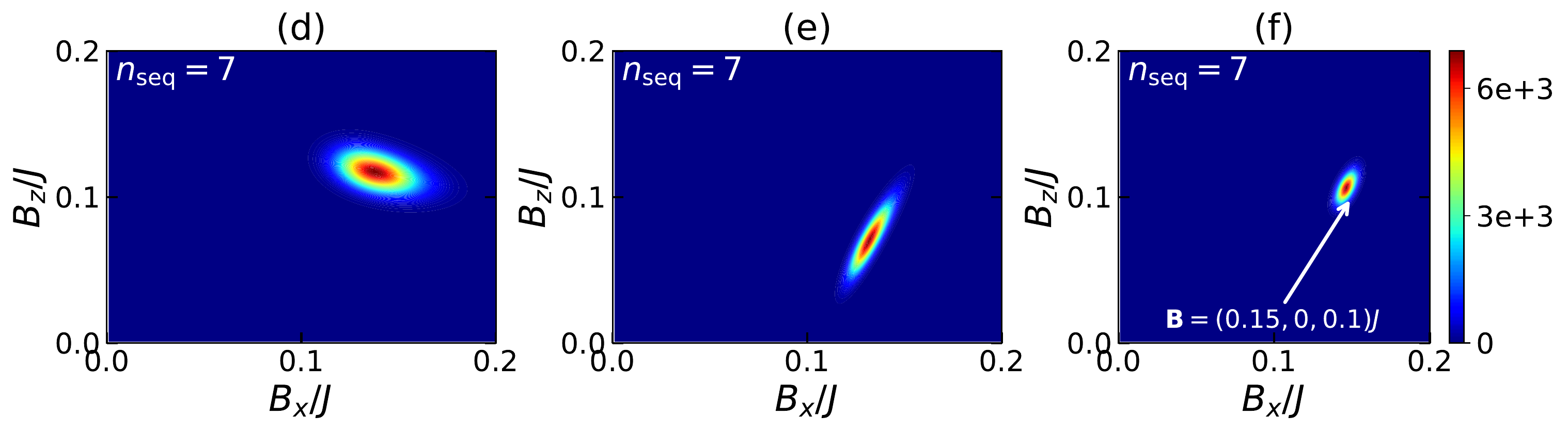}
\caption{Posterior distributions for estimating a true field $\pmb{B}=(0.15,0, 0.1)J$ with a spin chain probe of length $N=6$. Panels (a) to (c) computes the posterior distribution considering $n_\mathrm{seq} = 1$ and measurement basis $\sigma_N^x$, $\sigma_N^z$, and an overlapping of both, respectively. Panels (d) to (f) calculates the posterior distribution using $n_\mathrm{seq} = 7$ consecutive measurements with same measurement basis as before, i.e., $\sigma_N^x$, $\sigma_N^z$, and an overlapping of both, respectively. Our sequential sensing protocol shows a significant quantum-enhanced estimation of a multi-component local field, evidenced by the remarkably shrinking uncertainty of the overlapped posterior. The data gathering process is performed with the same total protocol time T for a fair comparison (see Eq.~\eqref{eq:total-time} in the main text).}
\label{fig:multiparemeter-SM}
\end{figure}

To exemplify the multi-parameter estimation, in Figs.~\ref{fig:multiparemeter-SM}, we plot the posterior distributions as functions of $B_x/J$ and $B_z/J$ for a probe of length $N=6$ and two different $n_\mathrm{seq}$ measurements when the true field is $\bm{B}=(0.15,0,0.1)J$. In Fig.~\ref{fig:multiparemeter-SM}(a), we consider $n_\mathrm{seq} = 1$ using $\sigma_N^x$ as measurement basis for $M=1000$ times. As seen from the figure, one can not completely infer the value of the multi-component magnetic field due to an emerging multi-valued posterior. This is because a single projective measurement entails only two outcomes, and therefore, for $n_\mathrm{seq} = 1$ there are several $B_x$ and $B_z$ with the same probability distribution satisfying the observed data. In Fig.~\ref{fig:multiparemeter-SM}(b), we consider $n_\mathrm{seq} = 1$ using this time $\sigma_N^z$ as measurement basis for another $M=1000$ trials. The same behavior can be observed in this case. Since the unknown local magnetic field is fixed, one can expect that the unknown field belongs to the intersection between the posteriors previously shown in Figs.~\ref{fig:multiparemeter-SM}(a)-(b). A simple overlapping of these probability distributions (which is equivalent to using the above set of POVMs) is shown in Fig.~\ref{fig:multiparemeter-SM}(c), where the posterior had been shrunk over $B_x/J$ and $B_z/J$ region. To show the advantage of our sequential sensing protocol, in Fig.~\ref{fig:multiparemeter-SM}(d), we consider $n_\mathrm{seq} = 7$ consecutive measurements using $\sigma_N^x$ as measurement basis (for the same execution time as for the $n_\mathrm{seq}=1$). Here, in contrast to the case $n_\mathrm{seq}=1$, due to the consecutive sequential measurements, the posterior gives a fair estimation even for a single measurement basis. In Fig.~\ref{fig:multiparemeter-SM}(e), we consider $n_\mathrm{seq} = 7$ using $\sigma_N^z$ as measurement basis. Very similar behavior is found as in Fig.~\ref{fig:multiparemeter-SM}(d). Finally, in Fig.~\ref{fig:multiparemeter-SM}(f), we overlap the posteriors, showing a notable reduction in the uncertainty of the unknown magnetic field. This confirms the power of our sequential protocol, where quantum-enhanced magnetometry significantly surpasses the one obtained by a conventional strategy for the same considered resources.

\section{iv. Protocol Robustness}

\subsection{A. Sequential sensing scheme in the presence of decoherence}

\begin{figure}
\includegraphics[width=0.75\linewidth]{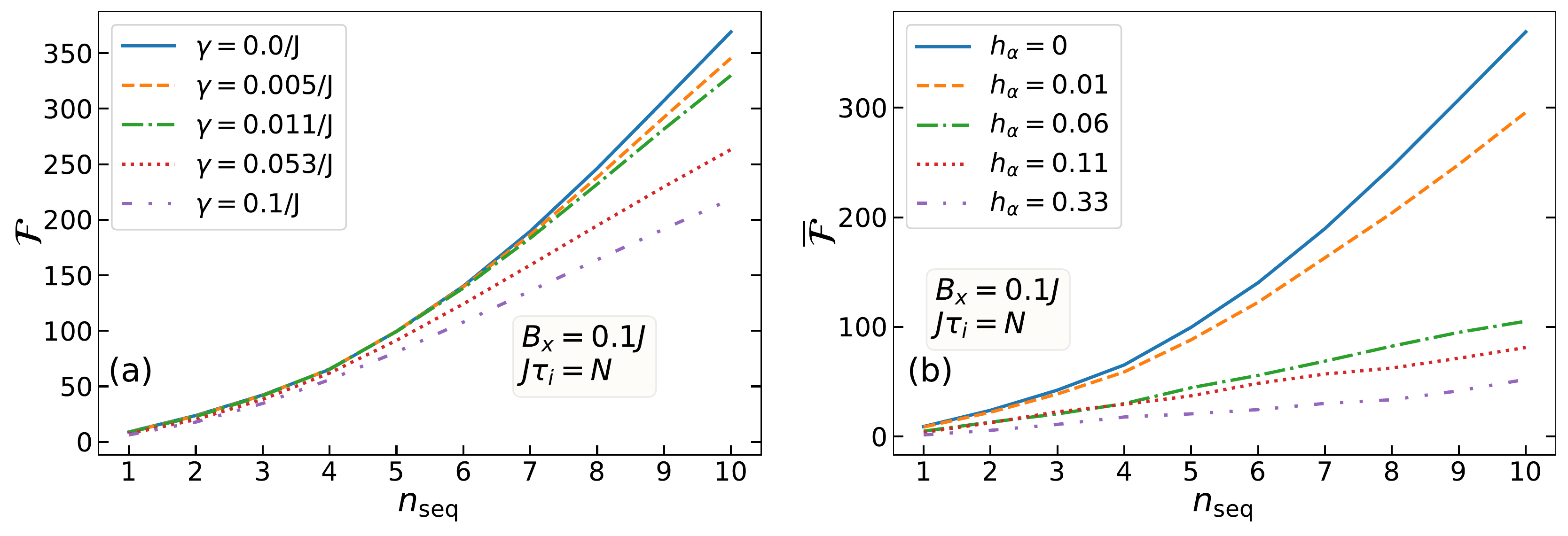}
\caption{(a) Classical Fisher information $\mathcal{F}$ as a function of the number of sequences $n_\mathrm{seq}$ performed on last site of a lossy system undergoing local dephasing at rate $\gamma$. (b) classical Fisher information $\overline{\mathcal{F}}$ averaged over 1000 samples as a function of $n_\mathrm{seq}$ where the system includes coupling anisotropies $J_\alpha \rightarrow J_\alpha + \Delta J_\alpha$ ($\forall \alpha = (x,y,z)$). Here, each $\Delta J_\alpha$ varies randomly and independently over the region $[-h_\alpha,h_\alpha]J_\alpha$. In both panels we aim to estimate the value $B_x=0.1J$ employing a system's probe of length $N=6$, performing sequential measurements at regular time intervals of $J\tau_i=J\tau=N$, and $J_\alpha = J$.}
\label{fig:revisions_robustness}
\end{figure}

Proposals for quantum-enhanced sensing must consider the possible noise and decoherence during the evolution (see Ref.~\cite{Garbe2021} in the main text). Indeed, these processes can not be avoided in real scenarios, and inspecting the proposed protocol's robustness against these incoherent processes is relevant. To determine up to which decoherence values our protocol can be accommodated, we consider the dynamics of each spin undergoing a local dephasing process at a rate $\gamma$. To do so, we model the dynamics with the following master equation
\begin{equation}
\dot{\rho} = -i[H, \rho] + \sum_{j=1}^N\frac{\gamma}{2}(\sigma^z_j\rho\sigma^z_j{-}\rho).
\end{equation}
In Fig.~\ref{fig:revisions_robustness}(a), we plot the classical Fisher information $\mathcal{F}$ as a function of the number of sequences $n_\mathrm{seq}$ for different values of $\gamma$. As the figure shows, even in the presence of strong dephasing rates $\gamma{\lesssim}0.05J$, the $\mathcal{F}$ still increases nonlinearly as $n_\mathrm{seq}$ increases. This could be understood as the measurement performed on the last spin reduces the entropy of the whole system, thus, entailing a purification mechanism at each measurement step. Since the local magnetic field $\boldsymbol{B}$ acts throughout the evolution, one can always extract some information regarding $\boldsymbol{B}$ by employing our protocol. The sequential measurement procedure then proves to be robust against dephasing, namely up to $\gamma \sim 5$\%$J$ , with the measurement steps aiding the system in overcoming it.

\subsection{B. Precision limits in the presence of anisotropies}

Another source of noise arising from the probe preparation is the inclusion of weak anisotropies (disordered couplings) in the exchange coupling $J$. From an experimental point of view, we consider such perturbations to be unknown and random every time the system is restarted. This means that, once a trajectory of $n_\mathrm{seq}$ measurements on the last site are completed, the protocol is reset; hence a new independent choice of random anisotropies for each $\alpha=(x,y,z)$ component takes place. We consider the following modified Hamiltonian with coupling anisotropies as:
\begin{equation}
H=-\sum_{\substack{j=1\\ \alpha=x,y,x}}^{N-1} (J_\alpha \pm \Delta J_\alpha)\sigma_j^\alpha \sigma_{j+1}^\alpha + B_x \sigma_1^x,\label{eq:Ham-anisotropies}
\end{equation}
where, in general, we consider $J_\alpha=J, \forall \alpha = (x,y,z)$, and the random and independent anisotropy strength varying over an interval as
\begin{equation}
\Delta J_\alpha \in [-h_\alpha,h_\alpha]J_\alpha.\label{eq:js}
\end{equation}
To set the precision limits of $B_x$ in the presence of anisotropy, we compute the classical Fisher information averaged over 1000 samples $\overline{\mathcal{F}}$. In Fig.~\ref{fig:revisions_robustness}(b) we plot the averaged classical Fisher information $\overline{\mathcal{F}}$ as a function of the number of sequences $n_\mathrm{seq}$ for different values of anisotropies. As the figure shows, the precision in estimating $B_x$ reduces significantly in the presence of this sort of noise. One could set a reasonable anisotropy tolerance at roughly 5\% of the exchange coupling between sites, i.e., $\Delta J \in [-0.05, 0.05]J$. Beyond this value, the $\overline{\mathcal{F}}$ losses its nonlinear features rapidly.

\subsection{C. Estimation limits in the presence of anisotropies}
\begin{figure}
\includegraphics[width=0.48\linewidth]{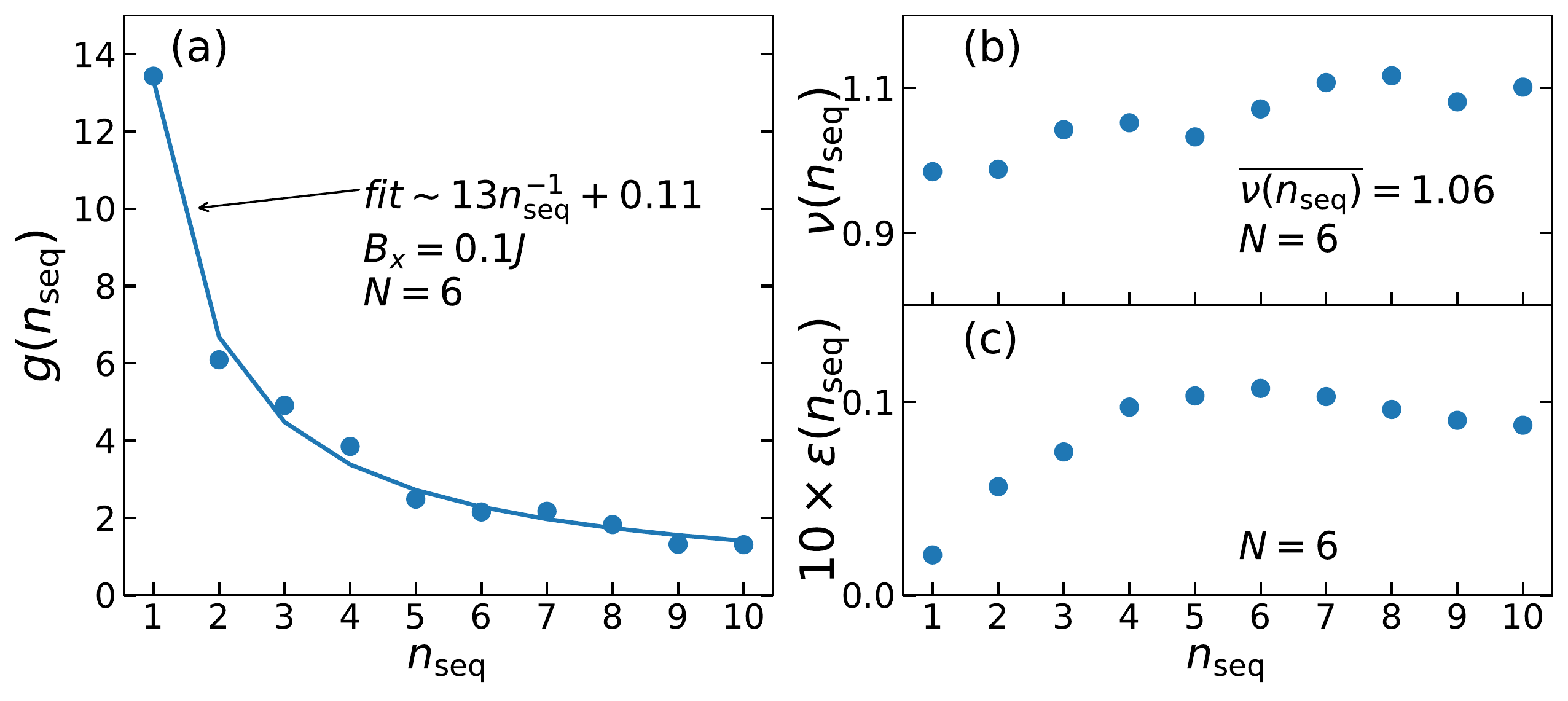}\includegraphics[width=0.452\linewidth]{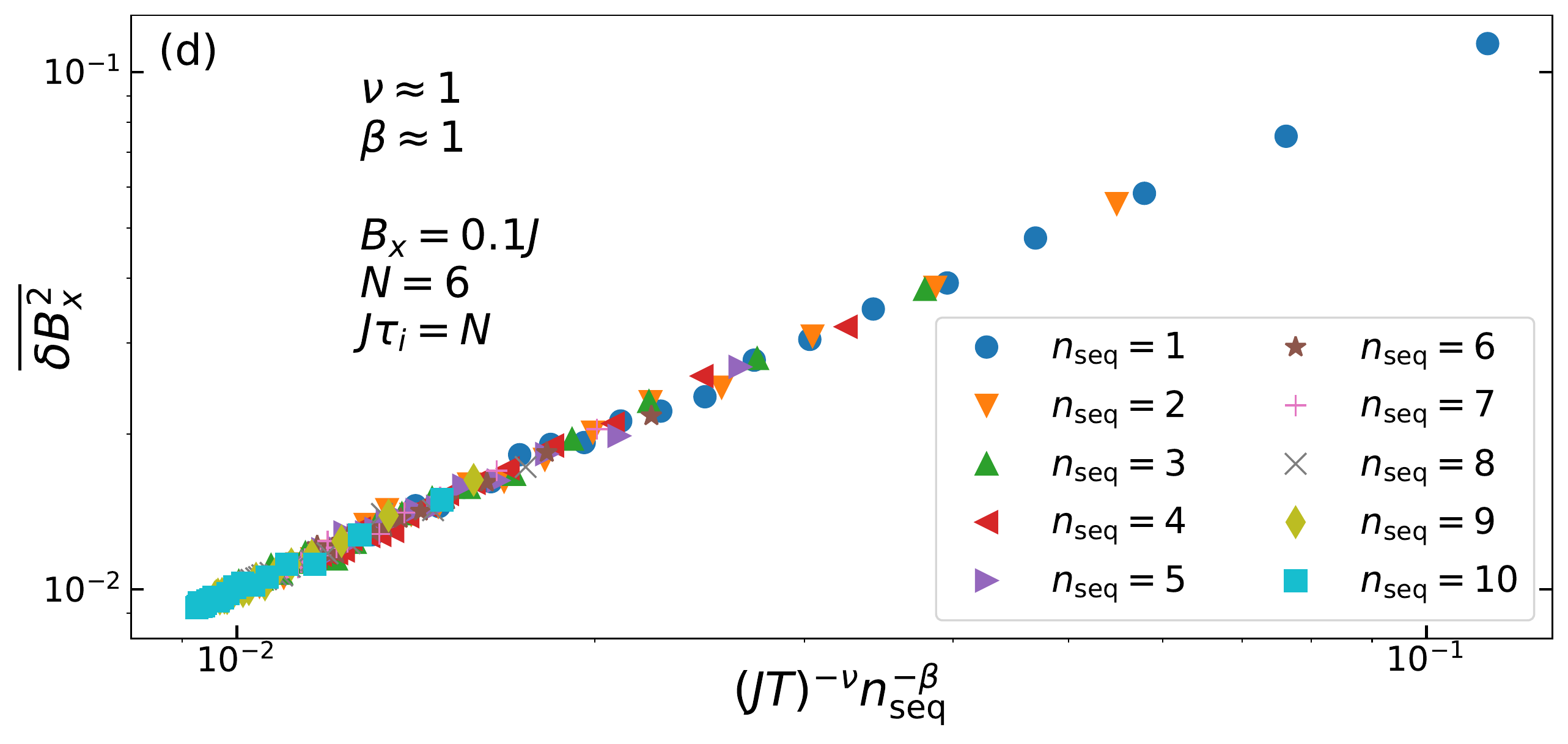}
\caption{Panels (a)-(c) illustrate the fitting coefficients from Eq.~\eqref{SM-fit} as a function $n_\mathrm{seq}$. In panel (a), $g(n_\mathrm{seq})$ presents a clear dependence on the number of sequential measurements that can be fitted using a polynomial function $g(n_\mathrm{seq}) \sim n_\mathrm{seq}^{-1}$. Panel (b) shows that the exponent $\nu(n_\mathrm{seq})$ fluctuates around an averaged value of $1.06$. In (c), the coefficient $\epsilon(n_\mathrm{seq}) \sim 10^{-2}$ does not play a relevant role in the final fitting form. Panel (d) plots the $\overline{\delta B_x^2}$ as a function of $(JT)^{-\nu}n_\mathrm{seq}^{-\beta}$ for the anisotropy case, here $h_\alpha = 0.01$, with excellent fitting behaviour. As the figure shows, even for this case one obtains quantum-enhancement in the estimation of $B_x$ as the number of measurement sequences increases.}
\label{fig:exponents-SM}
\end{figure}

As found in the main text, quantum-enhanced sensing via sequential measurements using time as a resource is shown to scale as
\begin{equation}
\overline{\delta B_x^2} \sim T^{-\nu} n_\mathrm{seq}^{-\beta}.
\end{equation}
\begin{figure}[t]
\includegraphics[width=0.32\linewidth]{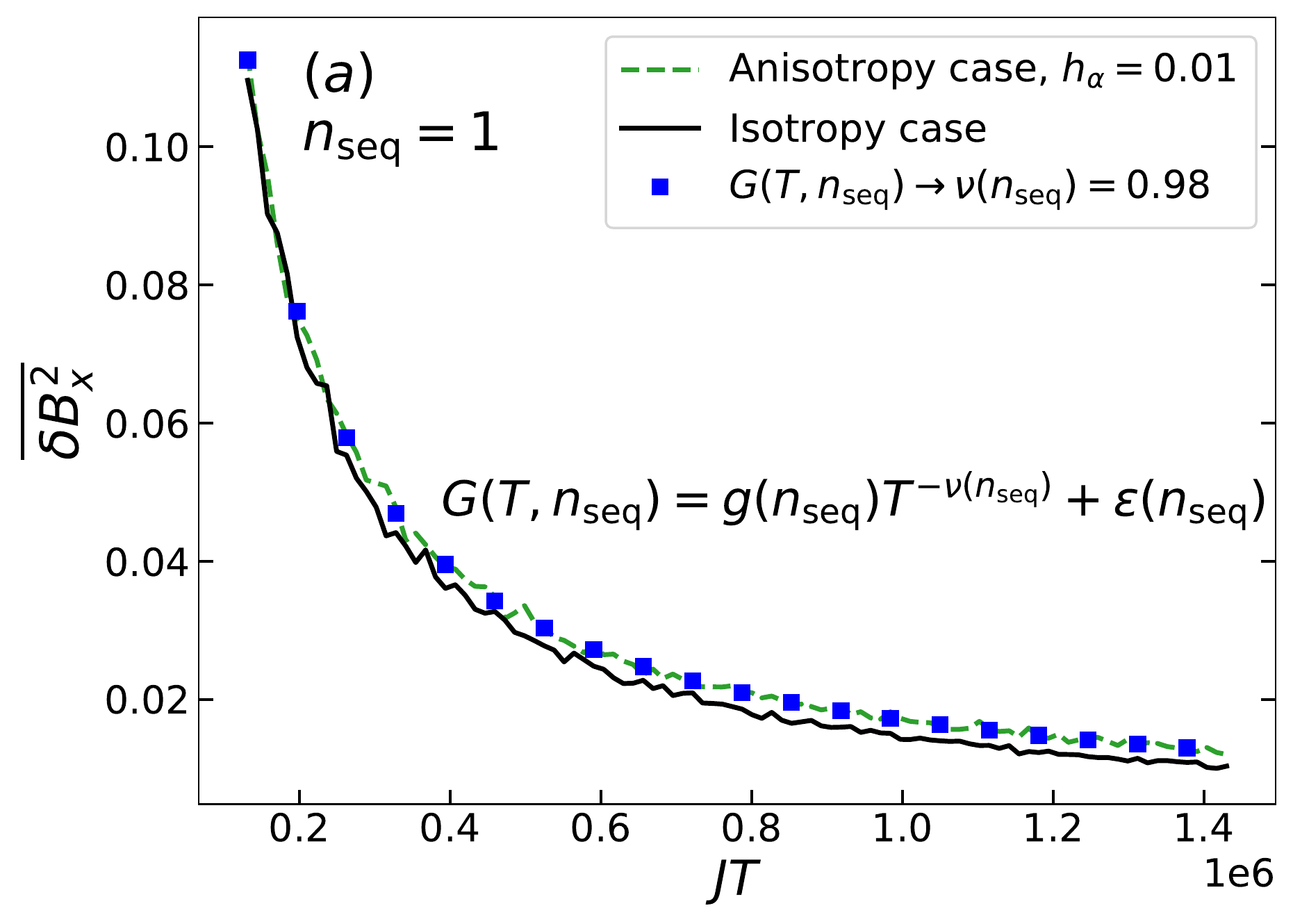}
\includegraphics[width=0.32\linewidth]{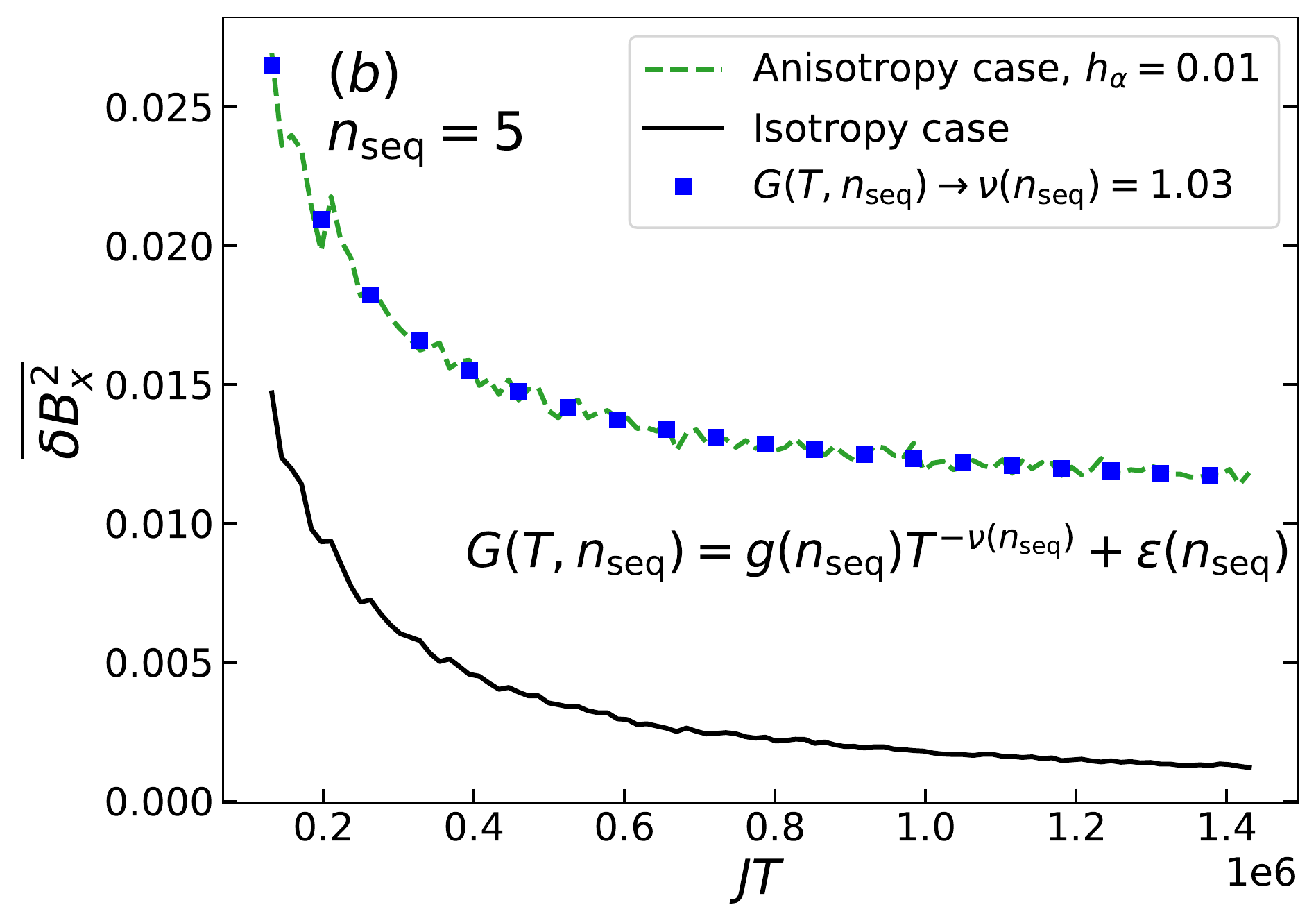}
\includegraphics[width=0.32\linewidth]{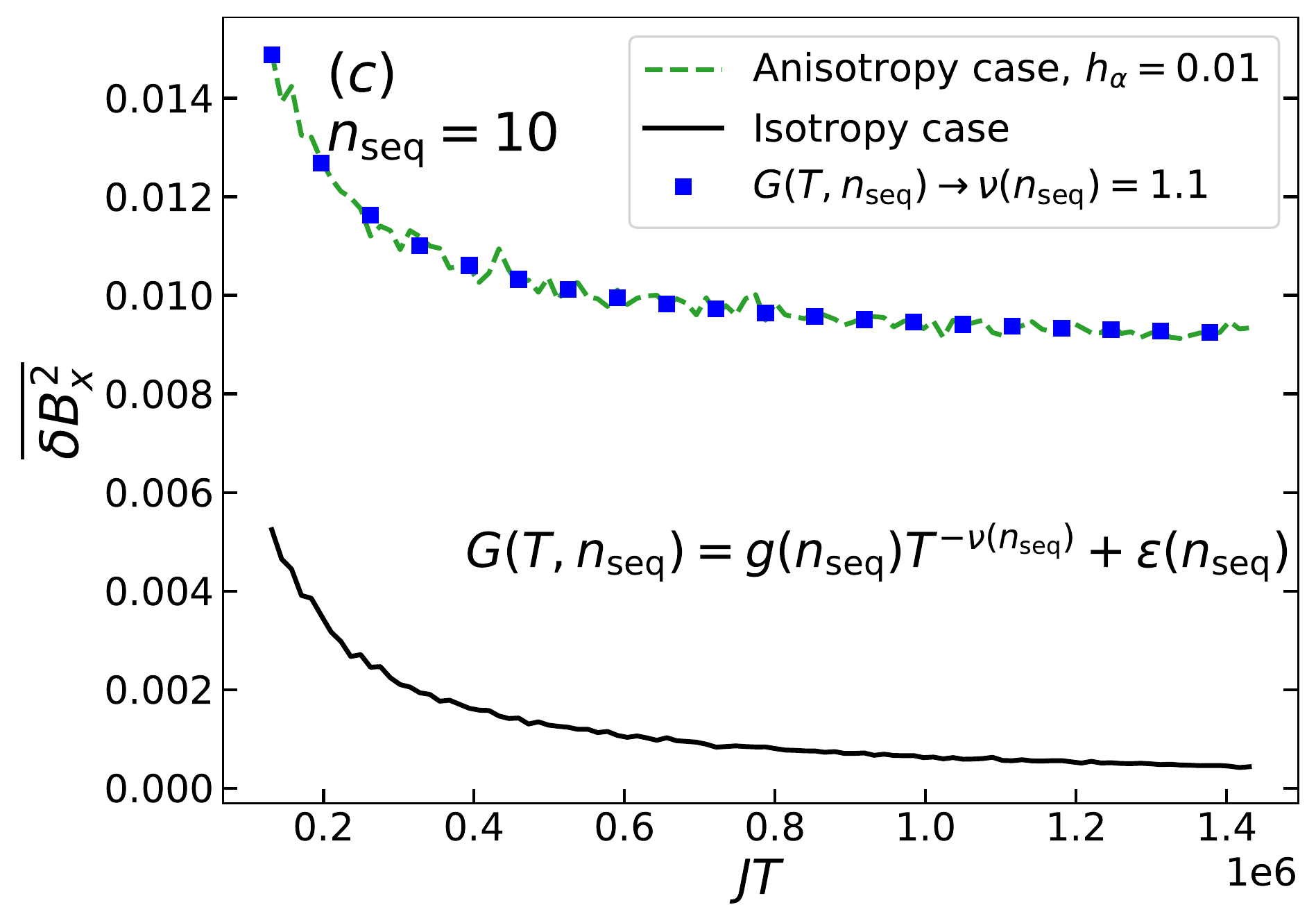}
\caption{Average squared relative error $\overline{\delta B_x^2}$ as a function of the total time $JT$ for different number of measurement sequences $n_\mathrm{seq}$. In each panel, we compare the $\overline{\delta B_x^2}$ when the parameter is estimated via Bayesian analysis from an isotropic (solid black line) or an anisotropic (dashed green line) case. In the case of random anisotropies (disordered couplings) scenario, we assume an isotropic model, whereas the observed data comes from a probe with coupling anisotropies, here $\Delta_\alpha = 0.01J_\alpha$, $\forall \alpha = (x,y,z)$. As seen from the figure, $\overline{\delta B_x^2}$ still behaves polynomially in the presence of anisotropies, with an exponent on average $\overline{\nu(n_\mathrm{seq})} \approx 1$, the rest of the fitting coefficients are analyzed in Fig.~\ref{fig:exponents-SM}.}
\label{fig:db2-ani-SM}
\end{figure}
In what follows, we attempt to extract the exponents $\nu$ and $\beta$ when the system is subjected to random coupling anisotropies. To make an estimation closer to experiments, we conduct Bayesian analysis by assuming the probability distributions $f(\pmb{\gamma}_j|B_x)$ coming from an isotropic probe modeled as in Eq.~\eqref{eq:hamiltonian-single}, whereas the observed data includes unknown random anisotropies varying over an interval $\Delta J_\alpha \in [-h_\alpha,h_\alpha]J_\alpha$, see Eq.~\eqref{eq:Ham-anisotropies}. Note that it is known that Bayesian analysis with a misspecified model undermines the performance of the estimator, and therefore, one should expect a decrease in the final local magnetic field estimation using this particular estimator in our protocol. In Figs.~\ref{fig:db2-ani-SM}(a)-(c), we compute $\overline{\delta B_x^2}$ as a function of the total time $JT$ for an anisotropy region of $h_\alpha = 0.01, \forall \alpha=(x,y,z)$ and for different choices of sequential measurements $n_\mathrm{seq}$. As the figure shows, the difference between the estimation of $\overline{\delta B_x^2}$ using a correct model (i.e., an isotropic model with observed data without anisotropies) becomes way more apparent as $n_\mathrm{seq}$ increases. Notably, regardless of the detrimental effects due to the anisotropies in the estimation, one still observes that $\overline{\delta B_x^2}$ decreases polynomially, and hence, one could fit a function of the form: 
\begin{equation}
G(T, n_\mathrm{seq}) = g(n_\mathrm{seq})T^{-\nu(n_\mathrm{seq})}+\epsilon(n_\mathrm{seq}).\label{SM-fit}
\end{equation}
In Figs.~\ref{fig:exponents-SM}(a)-(c), we plot the $n_\mathrm{seq}$-dependent coefficients extracted from Fig.~\ref{fig:db2-ani-SM}. As seen from Fig.~\ref{fig:exponents-SM}(a), the coefficient $g(n_\mathrm{seq})$ shows clear dependence on the number of sequences and it can be fitted with a polynomial curve $g(n_\mathrm{seq}) \sim n_\mathrm{seq}^{-1}$, whereas in Fig.~\ref{fig:exponents-SM}(b) the exponent $\nu(n_\mathrm{seq})$ fluctuates around an averaged value of $\overline{\nu(n_\mathrm{seq})}=1.06$. The coefficient $\epsilon(n_\mathrm{seq}) \approx 10^{-2}$ does not play a relevant role in the final fitting estimation. The above proper analysis, provides the final form for the average squared relative error
\begin{equation}
\overline{\delta B_x^2} \sim T^{-1}n_\mathrm{seq}^{-1}.
\end{equation}
From the above $\overline{\delta B_x^2}$ scaling, we conclude that our protocol can clearly accommodate up to disorder couplings within an interval of $h_\alpha = 0.01$, see Eq.~\eqref{eq:js}, and where the estimation is performed using misspecified Bayesian analysis. Remarkably, even for this case, one obtains quantum-enhancement in the estimation of $B_x$ with time as a resource as the number of sequences increases.

\end{document}